\documentclass{article}

\usepackage{capt-of,palatino}
\usepackage{authblk}

\usepackage[a4paper, total={6in, 9in}]{geometry}

\usepackage{color}

\usepackage[normalem]{ulem}
\usepackage{amsthm}

\usepackage{array}

\usepackage{hyperref}

\usepackage{algorithm}

\usepackage{algpseudocode}

\usepackage{algorithmicx}
\usepackage[textwidth=0.8in,textsize=tiny,color=yellow]{todonotes}
\usepackage{subcaption}
\DeclareCaptionSubType*{algorithm}

\DeclareCaptionLabelFormat{alglabel}{Alg.~#2}
\usepackage{todonotes} 

\usepackage[utf8]{inputenc} 
\usepackage[T1]{fontenc}    
\usepackage{hyperref}       
\usepackage{url}            
\usepackage{booktabs}       
\usepackage{amsfonts}       
\usepackage{microtype}      
\usepackage{epstopdf}
\usepackage{amsmath, amssymb}

\newtheorem{definition}{Definition}

\usepackage{lineno,xcolor}

\author[1]{Andreas Hula } 
\author[2]{Iris Vilares}
\author[3]{Peter Dayan}
\author[4]{P. Read Montague}

\affil[1]{\footnotesize{Wellcome Trust Centre for Neuroimaging, London, United Kingdom}}
\affil[2]{\footnotesize{Wellcome Trust Centre for Neuroimaging, London, United Kingdom}}
\affil[3]{\footnotesize{Gatsby Computational Unit, University College London, London, United Kingdom - Wellcome Trust Centre for Neuroimaging, London, United Kingdom}}
\affil[4]{\footnotesize{Wellcome Trust Centre for Neuroimaging , London, United Kingdom - Human Neuroimaging Laboratory, Virginia Tech Carilion Research Institute, Roanoke, VA, United States -  Department of Physics, Virginia Polytechnic Institute and State University, Blacksburg, VA, United States }}

\title{A Model of Risk and Mental State Shifts
  during Social Interaction} 

\begin{document}

\maketitle

\begin{abstract}
  Cooperation and competition between human players in repeated
  microeconomic games offer a powerful window onto social phenomena such
  as the establishment, breakdown and repair of trust.  However,
  although a suitable starting point for the quantitative analysis of
  such games exists, namely the Interactive Partially Observable Markov
  Decision Process (I-POMDP), computational considerations and
  structural limitations have hitherto limited its application, and left
  unmodelled some critical features of behavior in a canonical trust
  task. Here, we extend the I-POMDP framework and also improve
  inference. This allowed us to address two phenomena: a form of social
  risk-aversion exhibited by the player who is in control of the
  interaction in the game, and irritation or anger, seen as a shift of
  their internal state, exhibited by both players.  Irritation arises
  when partners apparently defect, and it potentially causes a
  precipitate breakdown in cooperation. Failing to model one's partner's
  propensity for it leads to substantial economic inefficiency. We
  illustrate these behaviours using evidence drawn from the play of
  large cohorts of healthy volunteers and patients.
  \end{abstract}

\begin{section}{Introduction}
  Assessing the internal characteristics of another person is a
  fundamental requirement for success in human social decision making.
  Neither people's self-reports, nor any current measurement device
  provides complete, veridical, information about another person's
  state. Nevertheless, we are typically quite adept at inferring the
  preferences and intentions of others and even at manipulating their
  states, in both cases over the course of multi-round interactions. One
  way to formalize this capacity is via the so-called interactive
  Partially Observable Markov Decision Process (I-POMDP; \cite{iPOMDP}).
  This is a regular Markov Decision Process (see \cite{Puterman})
  augmented with (a) partial observability (see \cite{Kaelbling}) about
  the characteristics of a partner; and (b) a notion of cognitive
  hierarchy (see \cite{Costa, Camerer}), associated with the game
  theoretic interaction between players who model each other.
  
  In recent work, we used approximate inference methods in the I-POMDP
  to capture the effect of an other-regarding utility preference (namely
  guilt) in modeling behaviour in a popular multi-round trust task (MRT)
  \cite{Brooks2008, Chiu, Debbs2008, Ting2012, Hula2015}. This model 
  offered powerful accounts of both the
  behavior of subjects, and also aspects of their neural activity
  \cite{Debbs2008, Ting2012}.
    
  However, a detailed inspection of the residual errors revealed two key
  characteristics that were missing from the model: social risk aversion
  and irritation. Here we formalize both, including extending the
  I-POMDP framework to encompass the possibility that players might
  change their internal states as a result of interactions. We thereby
  fit subjects' choices much more closely.

  First, investors are dominant in the MRT, in that they can still make
  substantial sums of money based on initial endowments in each round
  without investing anything. Perhaps as a result of this, we observed
  that some investors apparently treat a portion of their endowment as
  being exclusively theirs; only risking the remainder in the social
  exchange. This is a form of social preference that is absent in the
  Fehr-Schmidt model of other-regarding preferences that we adopted as
  our baseline model \cite{Fehr}. Here, we treat it explicitly as a form
  of (social) risk aversion, a factor that has previously been
  considered in terms of this task \cite{HouserTrustRisk}.

  A second, and more complicated, failure of the existing model is that
  sample investment profiles are generally too homogeneous. That is, as
  pointed out in some of the early neuroeconomics studies of the MRT
  \cite{Brooks2008}, cooperation between the players can readily break
  down in the face of apparent defection; with coaxing then being
  necessary to reestablish it (especially on the part of trustees). Such
  phenomena appear particularly prevalent in play involving subjects
  suffering from psychiatric conditions such as borderline personality
  disorder (BPD) (see for instance \cite{Brooks2008}).  This condition
  is frequently characterised by difficulties in maintaining social
  relationships, sudden ruptures in trust, and social withdrawal or
  aggression (see \cite{BPD, FonagyBPD}).

  We therefore augmented the model with a form of irritation. When
  irritated, subjects can exhibit substantially different rules of
  behaviour, for instance being unwilling to cooperate at all, and
  reducing their depth of interpersonal reasoning. This leads to
  breakdowns in cooperation. To predict what might happen in response
  to their own choices, and thus, if beneficial to them, to prevent a
  breakdown, subjects need to model the possibility of such a shift in
  their partner's state. They can then change their behavior
  prospectively. 

  As originally defined however, the I-POMDP framework explicitly
  excluded the possibility of one agent's actions changing the decision
  making preferences of the other agents (see p.$58$ \cite{iPOMDP}).
  This non-manipulability assumption is also in keeping with the
  conventional Bayes-Nash model \cite{Harsanyi}, in which nature
  allocates an agent's preference type before interactions start, and
  other agents merely make inferences about that type based on their
  observations.  We extended the I-POMDP framework to encompass the
  possibility of internal state shift manipulations, and indeed that
  other agents may be aware or unaware of the possibility of such shifts
  or the exact actions that might trigger them. This then gives rise to
  much richer dynamics and more intricate manipulations during social
  exchange.
  
  Our form of irritation fits somewhat better with the parameterized
  I-POMDP model of \cite{Wunder}. This allows players to have multiple
  potential policies or strategies about which their opponents make
  interactive inferences. However, although this framework can
  accommodate non-stationarity in the choice of strategy (players moving
  from being non-irritated to being irritated and vice-versa, as a
  result of their opponent's actions), it was originally designed with
  stationarity in mind. By contrast, our interpretation of a discrete
  internal state shift is critical for analyzing non-stationary human
  behaviour, in a manner that could for instance be used in functional
  magnetic resonance imaging (fMRI) analyses.

  We generated simulated data using our extended model to show how the
  inclusion of these dimensions of social manipulation affects the
  course and understanding of human social exchange, and to validate
  parameter identifiability. We then demonstrated how the new
  mechanisms allow us to account for behaviour that appeared anomalous
  according to our previous model. Finally, we discuss how this can further 
  the study of BPD and other disorders.

\subsection{Trust Task}
The multi round trust task (see \cite{Brooks2005, Ting2012, Hula2015},
based on \cite{McCabe}) (see figure \ref{fig:trusttask}A) is a
paradigmatic social exchange game. It involves two people, one playing
the role of an ``investor'' the other that of a ``trustee'', over $10$
sequential rounds. 
Quantities
pertaining to the investor and trustee are denoted by superscripts
``I'' and ``T'' respectively. 
 The participants played at the same time but did not know or meet each other 
 at any point.
 
 Both players know all the rules of the game. In each round, the investor
receives an initial endowment of $20$ monetary units. The investor can
send any $a^I$ units of this amount to the trustee. The experimenter
triples this quantity and then the trustee decides how much (an amount
$a^T$) to send back to the investor. This amount must be between 0
points and the whole amount that she receives. The repayment by the
trustee is not increased by the experimenter. After the trustee's
action, the investor is informed, and the next round starts. On each
round the financial payoffs of the two actors can be calculated: for the
investor this is:
\begin{align}
\chi^I (a^I, a^T) &= 20 -  a^I +  a^T 
\intertext{and for the trustee:}
 \chi^T(a^I, a^T) &= 3   a^I -   a^T.
\end{align}
For computational simplicity, the model treated the possible choices
on a coarser grid, allowing for five investor actions and five
corresponding trustee reactions.  The five investor actions correspond
to investing $0$, $5$, $10$, $15$ or $20$ or $\{0, \frac 1 4, \frac 1
2, \frac 3 4, 1\}$ of their endowment, while the trustee responses
correspond to the return of $0$, $\frac 1 6$, $\frac 1 3$, $\frac 1 2$
or $\frac 2 3$ of the received amount. The case in which the investor
gives $0$ is special, since the trustee has no choice but to return
$0$.  We round real subject actions to the respective nearest grid
point.

The Nash equilibrium (based on pure monetary outcomes) for this game
mandates a trivial interaction. That is because, in the last round, the investor should
never invest anything, since the trustee could defect without
punishment. Thus the interaction progressively unravels.  Real subject
behaviour in the game is quite different, and typically leads to
substantial investments and returns.
 
 \subsection{Generative Model}

A generative model of the multi round trust task was introduced in
\cite{Hula2015}; we enrich it here. Those of the parameters that we also
assume subjects to infer about each other over the course of interaction
are called ``intentional''; the other parameters are inferred by the
experimenter through the process of fitting the choices (using maximum
likelihood), but are merely assumed by the subjects and are constant
throughout the experiment.  Full details of the model can be found in
the supplementary material (in section \ref{OldModel}) here and in \cite{Hula2015}.
\begin{center}
\includegraphics[width=3.5in, height = 1.5in]{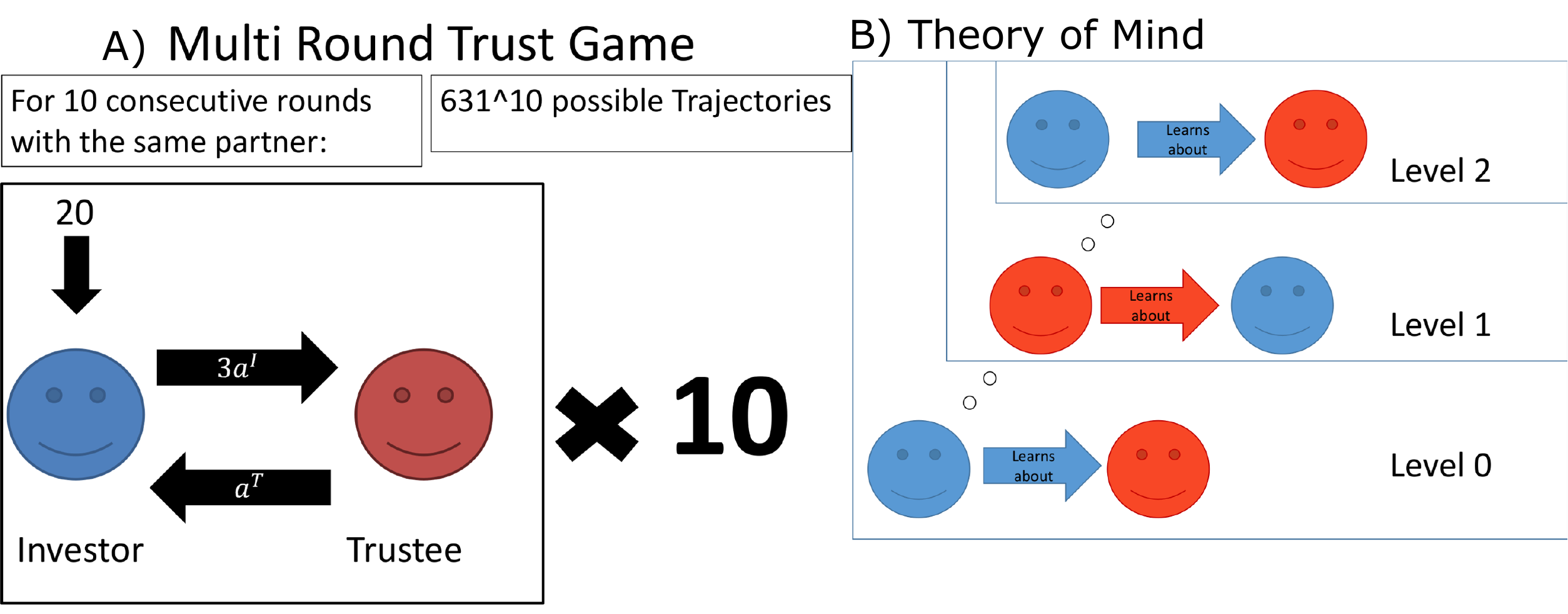}
\includegraphics[width=4.4in, height = 4.4in]{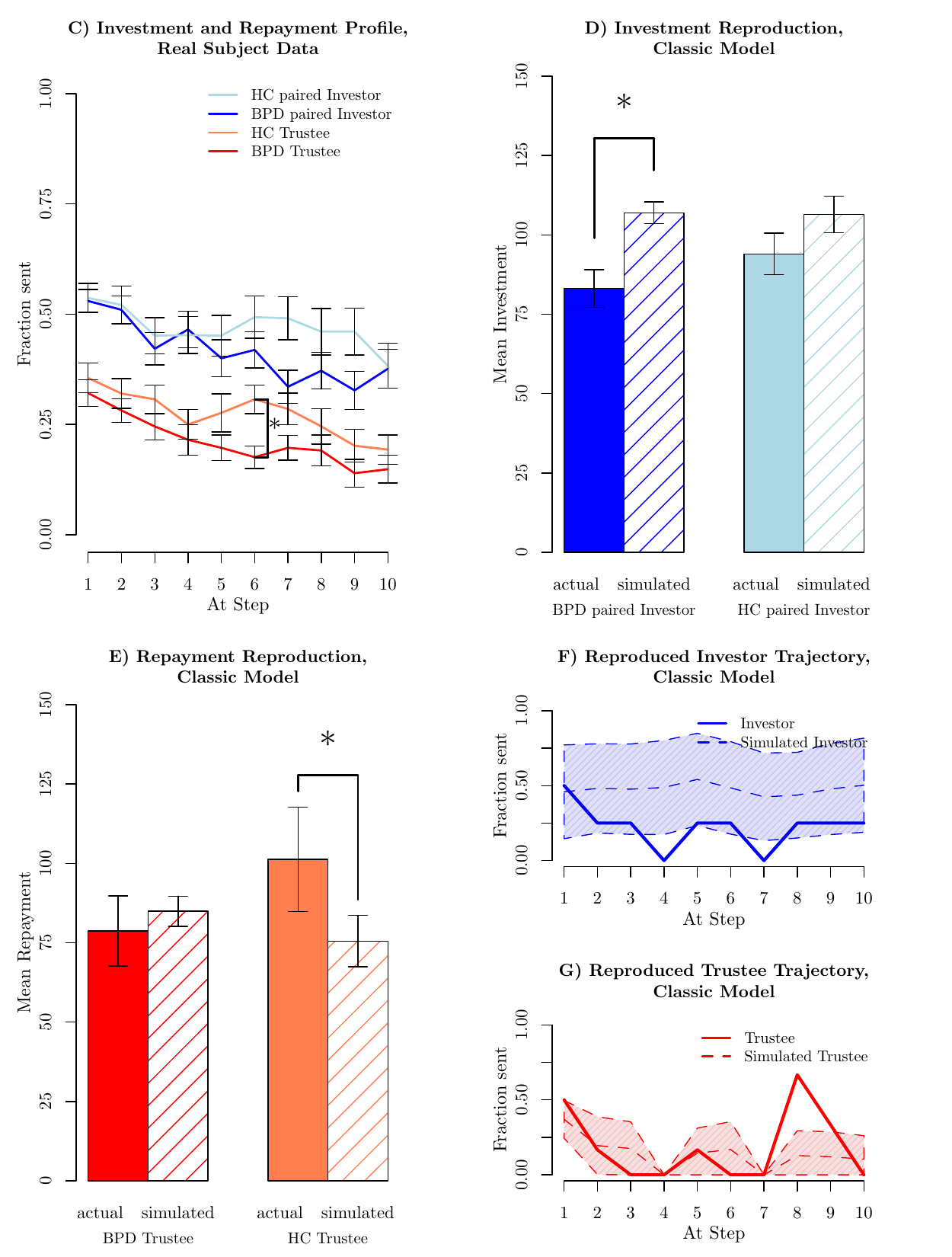}
	\captionof{figure}{\small{A: Physical features of the multi round trust game. 
	 B: Recursive reasoning about a
            partner. At level 0 the blue player learns about the partner.  
            At level 1 the blue players knows that the red player learns
            about them too (that is, that the red player is level 0).
            At level 2 the blue player knows that the red player knows
            they are learning about them (i.e. that the red player is
            level 1 and thinks of the blue player as level 0). This
            recurses up to higher levels. 
            C) Averaged investments and repayments in
    the data set. Errorbars show standard errors of the mean.  An asterisk denotes a significant difference
    ($p<0.05$, two sided t-test) corrected for multiple comparisons at the $10$ steps. D)
    Average investment in real and in simulated exchanges based on best
    fit parameters. An asterisk denotes a significant difference
    ($p<0.05$, two sided t-test) in means between the original data and
    the generated exchanges. E) Average repayments in real and in
    simulated exchanges based on the actual parameters. An asterisk
    denotes a significant difference ($p<0.05$, two sided t-test) in
    means between the original data and the generated exchanges.  F)
    Sample trajectory for an investor vs average of $200$ generated
    exchanges with best fitting parameters, based on the model in
    \cite{Hula2015}.  Shaded area shows estimated standard deviations.
    G) Sample trajectory for a trustee vs average of $200$ generated
    exchanges with best fitting parameters, based on the model in
    \cite{Hula2015}.  Shaded area shows estimated standard deviations.}
	\label{fig:trusttask}}
\end{center}

In the original model, there was a single intentional parameter,
namely guilt $\alpha\in\{0,0.4,1.0\}$. This denotes the sensitivity of
one player to inequality in their favour, as in the Fehr-Schmidt model
of inequity aversion \cite{Fehr}.  Subjects were assumed to use
Bayesian inference to infer their partners' guilt over the course of
the interaction. This is possible since a high guilt ($\alpha=1$)
partner will provide high investments or returns and appear
persistently cooperative, while a low guilt ($\alpha=0$) partner is
likely just to maximise their own winnings (and so only cooperate for
Machiavellian reasons).

Next, a player could be aware that their partner was also learning
about them, a recursive concept formalized as computational theory of
mind (ToM) or reasoning level $k$, and depicted in figure
~\ref{fig:trusttask}B.  A level $0$ investor learns about the trustee, but
treats her as being random rather than intentional. A level $1$
investor treats the trustee as being level $0$, implying that the
trustee is assumed to learn about a non-intentional investor. A level
$2$ investor treats the trustee as being level $1$, implying that the
trustee is assumed to know that the investor is learning about them
too. This continues recursively. One consequence of the interplay of
I-POMDP modeling and the asymmetric nature of the game is that only
even levels yield new insight into investor behaviour, and only odd
levels into that of the trustee \cite{Hula2015}. In the original
model, computational considerations restricted the theory of mind to
$k^I\in\{0,2 \}$ for the investor and $k^T\in\{0, 1 \}$ for the
trustee. In the MRT, levels of ToM higher than $4$ of ToM do not
appear to yield qualitatively new behavioural patterns (see
supplementary material \ref{OldToM}), and so we extended consideration to levels
$\{0, 2, 4 \}$ for investors and $\{0, 1, 3 \}$ for trustees.


Finally, subjects were classified according to their planning capacity
$P$, which quantifies how many steps of the future of the interaction
they take into account when assessing the consequences of their actions.
In the original model, this could take the values $P\in \{0, 2, 7 \}$.
However, it turns out that play for $P=4$ has very similar features to
that of $P=7$, involving exploitation of the partner and inhomogeneous
effects caused by the horizon of the game (see supplementary material \ref{OldPlan}).
Therefore, to liberate the computational capacity to model an additional
intentional parameter, we restricted $P$ to $\{ 1,2,3,4 \}$.

Choices were assumed to be made through a stochastic policy $\pi$ on the
basis of expected value action values (calculated in the manner of
\cite{Bellman}) through the medium of a logistic softmax (see
\cite{Centipede, 2step, nogo}). The inverse temperature parameter of the
softmax that was fixed at $\beta = \frac 1 3$ in the original model, was
here fit using values $\beta\in\{\frac 1 4, \frac 1 3, \frac 1 2,
1\}$. Note the relatively large numerical values of investment and
return, which is why the inverse temperatures may seem relatively
small compared with other studies.

To gauge the differences between models with different numbers of
parameters, we used the Bayesian Information criterion (BIC), which
penalizes the number of parameters $n$ used to fit each subject
according to the number $m$ of data points obtained in each exchange. It
is defined using the negative loglikelihood (NLL) $-\log \mathbb P[x_s |
\theta^*_s; M]$ at the best fitting parameters $\theta^*_s$ for each
subject $s$ under the given model $M$.
\begin{equation}
  \textrm{BIC}(M) = \sum_{\textrm{subjects
      $s$}}\left(-2\log \mathbb P[x_s | \theta^*_s; M]
+ n (\log (m) -\log(2\pi) \right)
 \end{equation} 
 In the multi round trustgame $m=10$, due to the $10$ choices per
 subject. The correction factor is for small $m$ \cite{Draper}.

The parameters of the final model can be seen in table ~\ref{tab:Par}.
\begin{center}
{\bf Table of Parameters}
\begin{tabular}{ | m{3cm} | m{3cm} |  m{8cm} | }
 \hline
Parameter & Values & Concept \\ 
 Guilt $\alpha $ & $\{ 0, 0.4, 1 \}$ & Measure of tendency to try 
 and reach a fair outcome.\\  
Plan $P$ & $\{1, 2, 3, 4 \}$ & Number of steps likely planned ahead.\\
Theory of Mind $k$ & $\{0, 2, 4 \}$ or $\{0, 1, 3 \}$ &  number mentalisation steps.\\    
Inverse Temperature $\beta$  & $\{1, \frac 1 2, \frac 1 3, \frac 1 4 \}$ & Certainty of own choice preference.\\    
 Risk Aversion (Belief) $\omega$ ($b(\omega)$)  & $\{0.4, 0.6, 0.8, 1.0, $ $1.2, 1.4, 1.6, 1.8 \}$ & Value of money 
 kept over (potential) money gained.\\    
  Irritability $\zeta$ & $\{0, 0.25, 0.5, 0.75,$ $ 1.0 \}$ & Tendency to retaliate on worse 
  than expected partner actions. \\    
 Irritation Belief $q(\zeta)$ & $\{0, 1, 2, 3, 4 \}$ & Initial belief on likelihood 
 of the partner being irritable.     \\  
  \hline
\end{tabular}
	\captionof{table}{\small{  All Parameters in the full model. }
	\label{tab:Par}}
\end{center}
\end{section}

\begin{section}{Materials and Methods}

\subsection{Ethics Statement}

Informed consent was obtained for all research involving
human participants, and all clinical investigation was conducted
according to the principles expressed in the Declaration of
Helsinki. The procedures were approved by the Institutional Board 
of Baylor College of Medicine.

\subsection{Subject Data}
We use the data set shown in \cite{Brooks2008}, consisting of $93$
healthy investors, paired with $93$ trustees, of which $55$ were BPD
diagnosed trustees (BPD Group, ''BPD'') and $38$ were healthy
trustees, matched in age, gender, IQ and socio-economic status (SES)
with the BPD trustee group (healthy control group, ''HC''). The
precise demographics can be found in \cite{Brooks2008}.

\subsection{Technical Data}

Programs were run at the local Wellcome Trust Center for
Neuroimaging (WTCN) cluster using Intel Xeon E312xx (Sandy Bridge)
processor cores clocked at $2.2$ GHz; no process used more than
$1.6$ GB of RAM.  We used R \cite{RStat} and Matlab \cite{MATLAB:2010}
for data analysis and the boost C++ libraries \cite{BoostSite} for code
generation. 

\subsection{Algorithmic Change}

The approach in \cite{Hula2015} utilized a sampling based method to
explore the decision tree during planning in the trust game, drawing
from approximate solution methods for tree search from machine learning
(see \cite{Auer1, Auer2} \cite{Szespari}\cite{POMCP2010}).  However,
if lower levels of calculation are all part of the same hierarchy, as well as kept in memory 
and so are immediately
available for higher level calculations, then the problem scales
linearly in the theory of mind level parameter, rather than
exponentially (as for other computational approaches to I-POMDPs
\cite{DoshiSample}, p.  $325$, $9.2.$). This trade off of memory for
computation is only practical if the planning horizon is reduced to at
most $4$ steps into the future. A more detailed discussion of the 
used algorithm can be found in the supplementary material 
\ref{Algo}.

The net result is that it takes less than $2$ minutes per generated 
$10$ step interaction, to calculate
deterministically (i.e., avoiding approximations from the stochasticity
of Monte Carlo-based tree evaluation) a $10$ step exchange of a level
$k^I=4$ investor with a level $k^T=3$ trustee, both having horizons of
$P=4$ steps. This comes at the cost of having to commit $0.8$ Gb of RAM
to the tree calculation.

\end{section}

\begin{section}{Results}

We start by illustrating the failures of the existing model of the
task. These motivate the changes that we then describe.

\subsection{Model Failure}

Figure ~\ref{fig:trusttask}C shows the average investments and returns in
the data from \cite{Brooks2008}. The dark blue and dark red lines in
figure~\ref{fig:trusttask}C show the respective average investments and
returns for healthy investors playing BPD trustees.  The lighter blue
and red lines show average investments and returns for healthy investors
and healthy trustees who matched the BPD trustees in socio-economic
status (SES), IQ, age and gender.  Investments averaged about half the
initial endowment and evolved over trials. In the second half of the
game, investors paired with BPDs invested considerably less than
investors paired with healthy trustees. This effect was a central topic
in \cite{Brooks2008}, and was explained by BPD trustees not heeding
warning signals from their investor partners, indicating investor
dissatisfaction with the BPD patients' lack of reciprocation.  The
significant difference ($p<0.05$, two sided t-test, Bonferroni corrected
for $10$ time step comparisons, indicated by an asterisk in figure
~\ref{fig:trusttask}C in trustee reciprocation at step $6$ also indicates
the time at which the average investment trajectories have persistently
diverged. This gave rise to the difference in early vs late investment
between the two groups that was reported in \cite{Brooks2008}.

The solid bars in figure~\ref{fig:trusttask}D show the average total
investments in the real data for the two groups. These differ
significantly at $p<0.05$ in a two sided t-test, as reported in
\cite{Brooks2008}. The hatched bars show the result of generating data from the
model in \cite{Hula2015} (using the extensions discussed above to higher
theory of mind and lower maximal planning). Model data is generated for
each dyad, using that dyad's best fitting parameters.  The model
overestimates the investments of the BPD-paired investors by
about $40\%$

Figure ~\ref{fig:trusttask}E demonstrates a similar issue for the modelled
trustees. The solid bars show the returns of the control and BPD
trustees; these again differ significantly at $p < 0.05$. Further, the
simulated HC trustees (hatched bars) return significantly less than the
actual HC trustees.  Although it may seem that the simulated BPD
trustees return similar proportions to the actual BPD trustees, this
actually flatters the model, since this repayment is based on the
over-generous model investment (the hatched bars in part B) rather than
the true, more miserly, investment.

A second model failure concerns the detailed dynamics of investment
across the task. The solid lines in figure~\ref{fig:trusttask}F;G show a
selected sample interaction between a healthy investor
(figure~\ref{fig:trusttask}F) and a BPD trustee (figure~\ref{fig:trusttask}G).
The trustee provides a poor return in trial $3$, and is met by zero
investment in trial $4$. The same pattern repeats in trials $6$ and $7$.
The trustee is then far more generous in trial $8$; this then coaxes (to
adopt a term from \cite{Brooks2008}) the investor to continue investing,
though after $2$ breaks, the investors is unwilling to much increase
their investment above a low level.  The trustee then defects on trial
$10$, returning nothing.  \cite{Brooks2008}'s conclusion was that a
significant portion of the BPD group lacked mechanisms that could
consistently repair the faltering interactions that occur when subjects
become what we will describe as being irritated. Thus tentative ruptures
(in the form of drops in investment level) turned into complete breaks,
with the investor using their position of power in the game to punish
the trustee.

The dashed lines in figure~\ref{fig:trusttask}F;G show the result of
simulating $200$ trajectories using parameters fit to the actual data,
and also making predictions at each step based on the actual investments
and returns of the dyad prior to each step (explaining why the model
return is also $0$ on trials $4$ and $7$). The shaded areas show the empirical standard deviations - which are
evidently very wide.  In fact, the specific reductions are not only
absent in the averages; the modelled investment following the trustee's
defection on trials $3$ and $6$ decreased to $0$ on only $11$\% and
$13.5$\% of the sample runs; compared with the collapse to $0$ apparent
in the actual data.

We addressed these sources of model failure by introducing two new
parameters, associated with risk aversion and irritation.

\subsection{Risk Aversion}

The investor is in charge in the MRT, since she could simply keep her
endowment on each round. It has been noted since the advent of this kind
of trust game in \cite{McCabe} that a lack of investment could represent
a social form of risk aversion rather than a lack of trust; see
\cite{HouserTrustRisk}. This could account for differences in levels of
investment regardless of the cooperativity of either partner.

We parameterize such risk aversion as a multiplicative factor $\omega^I$
in the payoff functions, increasing or decreasing the evaluation of
money that the investor keeps for herself compared to the money returned
by the trustee:
\begin{equation}
  \chi^I_{\omega} (a^I, a^T) = \omega^I (20 -  a^I) +  
a^T, 
\end{equation}
with $\omega^I \in [0.4,1.8]$ (in $7$ steps of $0.2$). The trustee is
subordinate in the task, and so does not have a risk parameter of their
own. Instead, the trustee makes an assumption about the investor's
degree of risk aversion, at one of the above mentioned $8$ values. We
capture intentional aspects of trust through guilt, and so treat risk
aversion as a non-intentional parameter. However, in keeping with
\cite{Harsanyi}, both players are assumed to be consistent, with the
investor believing the trustee to know her risk aversion, and to know
that she believes this; and the trustee believing that the investor
believes this too. We write $b^T(\omega^I)$ for the trustee's belief
about the investor's value of $\omega^I$.

Depending on the trustee's belief $b^T(\omega^I)$, there will be
either earlier or later attempts at exploitation. If $b^T(\omega^I) <
1$, then the trustee infers the investor will keep investing, even if
the trustee has been relatively uncooperative (i.e.  the investor will
be risk-seeking). Conversely, if $b^T(\omega^I) > 1$, then the trustee
will infer that any investment is contingent on their behavior, and
there could be negative consequences of poor return. For values
$b^T(\omega^I)> 1.4$, the trustee expects the investor to invest
so little that building up trust will not be worthwhile in the first
place. In this case, the interaction will rupture.

Illustrations can be found in the supplemental material \ref{RiskDetail},  
along with additional detail on the workings of this parameter.


Including risk aversion allows the model to account for the behavioural
data much more proficiently, with the average Investor NLL improving
from $12.94$ to $9.68$. The average trustee NLL improves from $11.36$ to
$9.5$. The average BIC for the investors improves from $27.3$ to
$21.68$, and for the trustees, from $23.98$ to
$21.3$.

\subsection{Irritation}

We explained the breakdown in cooperation evident in
figure~\ref{fig:trusttask}F;G as arising when the participants become
irritated.  Formalizing this leads to four considerations: (i) what do
subjects do differently when irritated; (ii) what leads a subject to
become irritated; (iii) how can irritation be repaired; (iv) and what do
subjects know about their own irritability?  We offer a highly
simplified characterization of all four of these. Individual
interactions in the $10$ round MRT are too short to license more complex 
treatments.

\begin{definition}[Irritability]
  We define the irritated state as associated with planning $P=0$, guilt
  $\alpha=0$, temperature $\beta = \frac 1 2$ and complete disregard of
  beliefs about the other player that have hitherto been established. Additionally, 
  for investors, the value of the risk aversion under irritation ($\omega^I_{\iota}$) is bounded 
  below at $1.0$ i.e. $\omega^I_{\iota} = \max \{1.0, \omega^I\}$, since 
  otherwise ``irritated'' investors may not show punishing behaviour.
  We model the players' policy $\pi$ as being a mixture between
  irritated $\pi_{\iota}$ and the nonirritated $\pi_{\tilde{\iota}}$
  choices, with irritation weight $v_{\iota}$
\[
\pi (a, h) =
(1-v_{\iota})\pi_{\tilde{\iota}} (a, h)+v_{\iota}\pi_{\iota}(a, h).
\]
A participant's irritation weight is assumed to start at
$v_{\iota}=0$, and to increase when their partner's
action (investment or return) falls short of the value expected on the
basis of the current model they have of the partner (including the 
partner's potential irritation):
\begin{align}
v_{\iota} &= \min\{ v_{\iota} + \zeta, 1.0\} \qquad \text{given unfavorable
  investment or return}
\intertext{where $\zeta$ is a subject-specific parameter. Irritation
  decreases through a process of repair when the  action exceeds this
  expected value} 
v_{\iota} &=  \max\{ v_{\iota} - \zeta, 0.0\} \qquad \text{given favorable
  investment or return}
\end{align}
\end{definition}

\begin{definition}[Intentional Inference about Irritation]
  Players maintain and update beliefs about the partner's irritability in
  exactly the same way as about the partner's guilt: that is, they
  employ a Dirichlet prior on a multinomial distribution over five
  possible irritation values $\zeta \in \{ 0, 0.25, 0.5, 0.75, 1\}$
  (dubbed respectively ``nonirritable'' and four different``irritable''
  types in the following) and use the same approximate inference rule as
  is used for guilt.
\end{definition}

However, unlike guilt, which we imagine is a characteristic that
varies continuously amongst our participants, we consider a discrete
set of possible prior beliefs about irritability. That is,
irritability awareness is treated as an additional discrete new
parameter ($q^I(\zeta^T);q^T(\zeta^I)\in\{0,1,2,3,4\}$).  The
investor's value $q^I(\zeta^T)$ determines prior weights of his belief
over the trustee's actual irritability $\zeta^T$. The trustee's value
$q^T(\zeta^I)$ determines prior weights of her belief over the
investor's actual irritability $\zeta^I$.  These priors are intended
to cover a suitable range of possibilities; as noted, the MRT involves
too few choices to license a richer depiction.
\begin{center}
{\bf Table of Irritation Inference settings}

\begin{tabular}{ |m{1cm} |m{0.1cm}  r@{.}l  m{0.5cm}  m{0.5cm} m{0.5cm}  r@{.}l  r|  m{2cm} | }
\hline
 \multicolumn{1}{|c|}{belief} & \multicolumn{9}{|c|}{ prior over $\zeta$} &  \multicolumn{1}{|c|}{descriptor} \\[2pt]\hline
$q(\zeta)$ & $\{$ & $0$&$0,$  & $ 0.25,$ & $0.5,$ & $ 0.75,$ &  $1$&$0$ & $ \}$
& \\[2pt]\hline
$0$ & $\{$ & $400$&$0,$ & $0.1,$ &  $ 0.1,$ &  $ 0.1,$ & $0$&$1$ & $ \}$ & ignorant \\[2pt]\hline
$1$ & $\{$ & $4$&$0,$ & $ 0.5,$ & $0.5,$ & $0.5$, &  $0$&$5$ & $ \}$ &
optimistic\\[2pt]\hline 
$2$ & $\{$ &  $0$&$4,$ & $0.1,$ & $0.1,$ & $0.1,$ & $0$&$1$ & $\}$ &
realistic\\[2pt]\hline 
$3$ & $\{$ & $2$&$0,$ & $ 1.0,$ & $ 1.0,$ &$ 1.0,$ &$1$&$0$ &$\}$ &
pessimistic\\[2pt]\hline 
$4$ & $\{$ & $ 0$&$1,$ & $ 0.1,$ & $0.1,$ & $0.1,$ & $400$&$0$ & $\}$ &
fatalistic\\[2pt]\hline 
\end{tabular}
	\captionof{table}{\small{Irritability belief settings.}
	\label{tab:irrbel}}
\end{center}

Table~\ref{tab:irrbel} lists the four particular prior beliefs
$q(\zeta)$ over the values of irritation. Players range from being
\emph{ignorant} about the possibility that their partners might be
irritable, through stages of \emph{optimism} that they are not,
\emph{realism} that they could be, \emph{pessimism} that they likely are
and \emph{fatalistic} that they certainly are.

Finally, although we assume that players infer both their partner's
inequality aversion and their partner's irritability level during the
interaction, we do not allow subjects to consider their \emph{own}
future irritation. This follows \cite{Loewenstein}'s observations of
subjects' inability whilst engaging in 'cold' cognition to contemplate
the possibility of 'hot' cognition.

A detailed example of the general workings of irritation in the case of
a single trajectory with potentially aware participants
($q^I(\zeta^T)=q^T(\zeta^I)=2$) is shown in figure \ref{fig:Brooks}A.
The golden line depicts the evolution of the irritation weight
$v_{\iota}^I$.  At step $2$, a subpar repayment by the trustee was
introduced by fiat to irritate the investor (the expected repayment by
the trustee would have been 50\%).  The investor's irritation duly rose
to $v^I_{\iota}=0.5$.  At this point the trustee's belief about the
investor's irritability is still at $0.5$, as they have not observed
the investor's response to their action. 
At step
$3$ the investor retaliated against the earlier defection of the
trustee. The aware trustee thus updated their irritation beliefs,
inferring that the investor was more likely to be irritable (at a
marginal probability of $p = 0.58$).  Noting the potential cost to the
interaction of further irritating the investor, the trustee ensured a
better than expected response in the next interaction at step $4$. Not
only did the trustee repair the interaction, they also ensured that
they did not further irritate the investor, at least until the very
end of the interaction, as can be seen in the remainder of figure
\ref{fig:Brooks}A, from step $4$. This exactly captures the
``coaxing''-type repair mechanism that \cite{Brooks2008} suggested to
explain differences in investment behaviours elicited by healthy
control and BPD trustees.

Figure \ref{fig:Brooks}B shows the consequence of a lack of
irritation inference in the presence of an irritable investor. The
players had the same parameter values as in figure \ref{fig:Brooks}A,
except for being irritability ignorant ($q^I(\zeta^T)=q^T(\zeta^I)=0$).
After the same two initial actions (again introduced by fiat), without a
notion of the partner being irritable, the trustee missed the chance to
repair the interaction at step $3$ and the investor's irritability
weight rose to $v^I_{\iota}=1$.  From this point on the investments
stayed low and the trustee did not placate the investor, thus receiving
only a paltry total income. Both players failed to extract anything like
the full return available from the experimenter.

Quantitative effects of irritability on the group level can be found in the supplementary 
material \ref{IrrQuant}.

Figures~\ref{fig:Brooks}C;D show that including these various
features removes the discrepancies between data generated from the full
model and the subject data.  There is no longer a significant difference
between generated and original investments or repayments. The complete
model predicts $44\%$ of the investor choices (chance is $20\%$) or
equivalently an average NLL of $8.41$ on $10$ investor choices (form
$9.68$) and an average NLL of $7.5$ or $47\%$ of choice predicted for
trustee choices (from $9.5$). The final average BIC for the investors is
$20.07$ and for the trustees is $18.2$.  

Figure~\ref{fig:Brooks}E;F demonstrates that the model qualitatively
captures ruptures and repair occurring in real interactions, with the
investment decreasing to $0$ on $50$\% of the sample runs on trials $4$
and $7$.  Further, the spread of the predictions is greatly reduced from
those in figure~\ref{fig:Brooks}E;F.  The investor NLL of this
interaction improves from $7.47$ to $5.42$, while the trustee improves
from NLL $11.53$ to an NLL
of $9.5$ (with $\zeta^I=0.5; \zeta^T=1$).

\begin{center}
\includegraphics[width=5.5in, height = 5.5in]{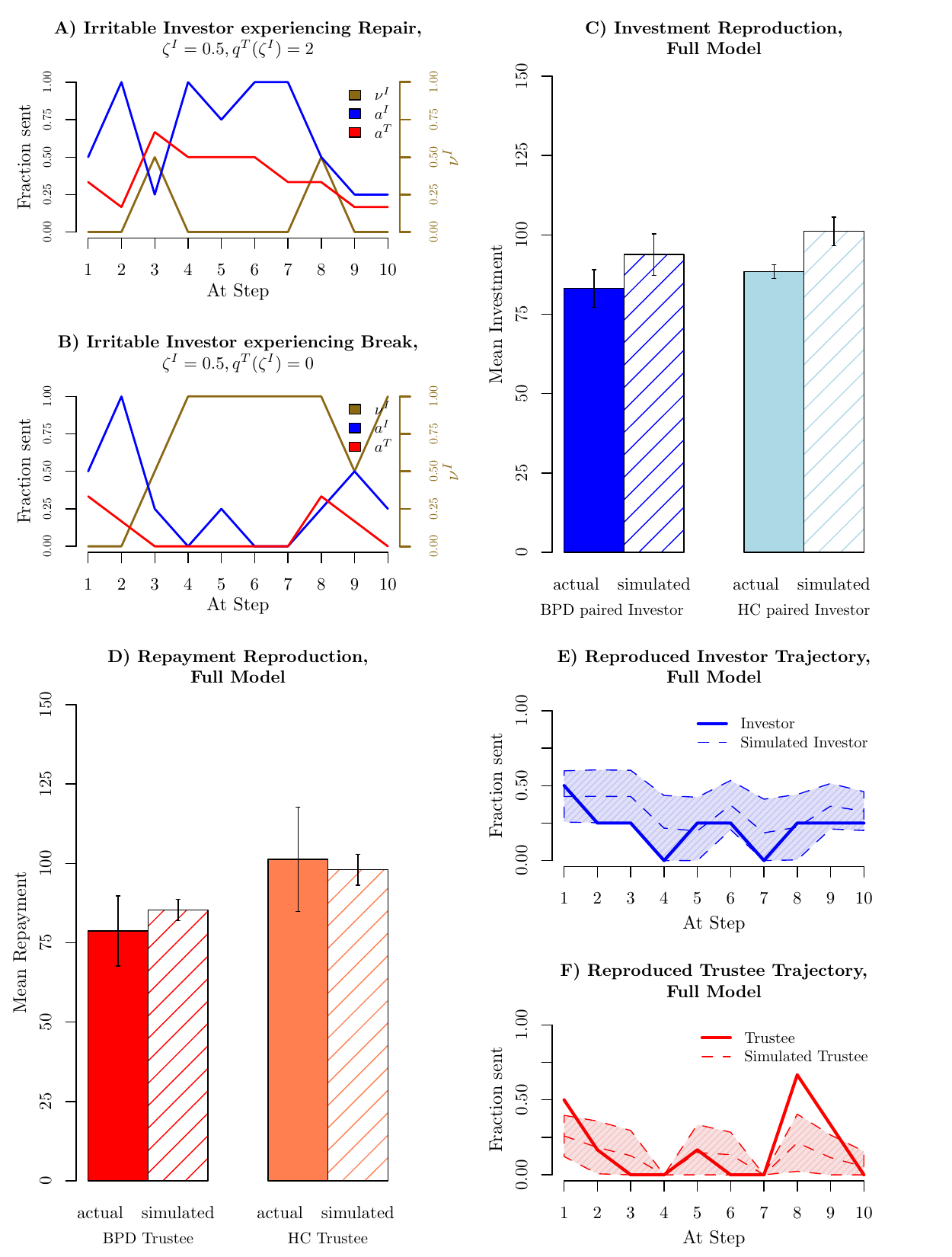} 

\captionof{figure}{\small{
    A) Simulated Repair Interaction. Single
    trajectory of two aware players (blue for investor, red for trustee). The golden line
    depicts the evolution of the investor irritation weight during the
    interaction. B) Simulated Break Interaction. Both players were
    irritability ignorant, thus they do not notice potential
    irritation. The gold line depicts the evolution of the investor
    irritation weight during the interaction (its value at the \emph{start} of the relevant round is shown). For A;B the simulated
    investor/trustee had $k=2/1, \zeta=0.5/0, \alpha=0.4, P=4,
    \beta=\frac 1 3$.
  C) Average Investment profiles
            regenerated from estimated parameters  
	 in the full model. All errorbars are standard error of the mean.
	D) Average Repayment profiles regenerated from estimated parameters 
	 in the full model. All errorbars are standard error of the mean.  E) Reproduction of sample investor trajectory using 
	 $200$ simulated interactions with the best fitting parameters. 
	 Shaded areas are estimated standard deviation.
  F) Reproduction of sample trustee trajectory using 
	 $200$ simulated interactions with the best fitting parameters. 
	 Shaded areas are estimated standard deviation.	   }
	\label{fig:Brooks}}
\end{center}
\subsection{Behavioural Analysis}

The main intent of refining the model was to use it to make inferences
about the two investor and two trustee groups that generated the
data. In the supplemental material \ref{ModelInversion}, we show that 
such inferences are legitimate in that the parameters are broadly
identifiable in self-generated data. 

The distributions of the new parameters (risk aversion, irritability,
awareness) across the groups are shown in figure ~\ref{fig:Stat}A-F.  We
extend a finding reported in \cite{Hula2015}, namely that even in the
extended model, the average guilt in BPD trustees is significantly ($p=
0.03$, $\alpha^T : 0.32 < 0.49$, uncorrected for multiple comparisons)
lower in BPD trustees compared to matched (in IQ and socio-economic status) 
healthy controls. This can be traced
back to a significantly($p=0.03$, $\chi^2$-test for equal proportions,
uncorrected for multiple comparisons) higher proportion of guilt
$\alpha^T=0$ subjects.  Additionally, the irritation ignorant awareness
setting ($q^T(\zeta)=0$) is significantly ($p=0.03$, $\chi^2$-test for
equal proportions, uncorrected for multiple comparisons) more common in
BPD trustees, compared to HC trustees.

 \begin{center}
\includegraphics[width=5in, height = 6.5in]{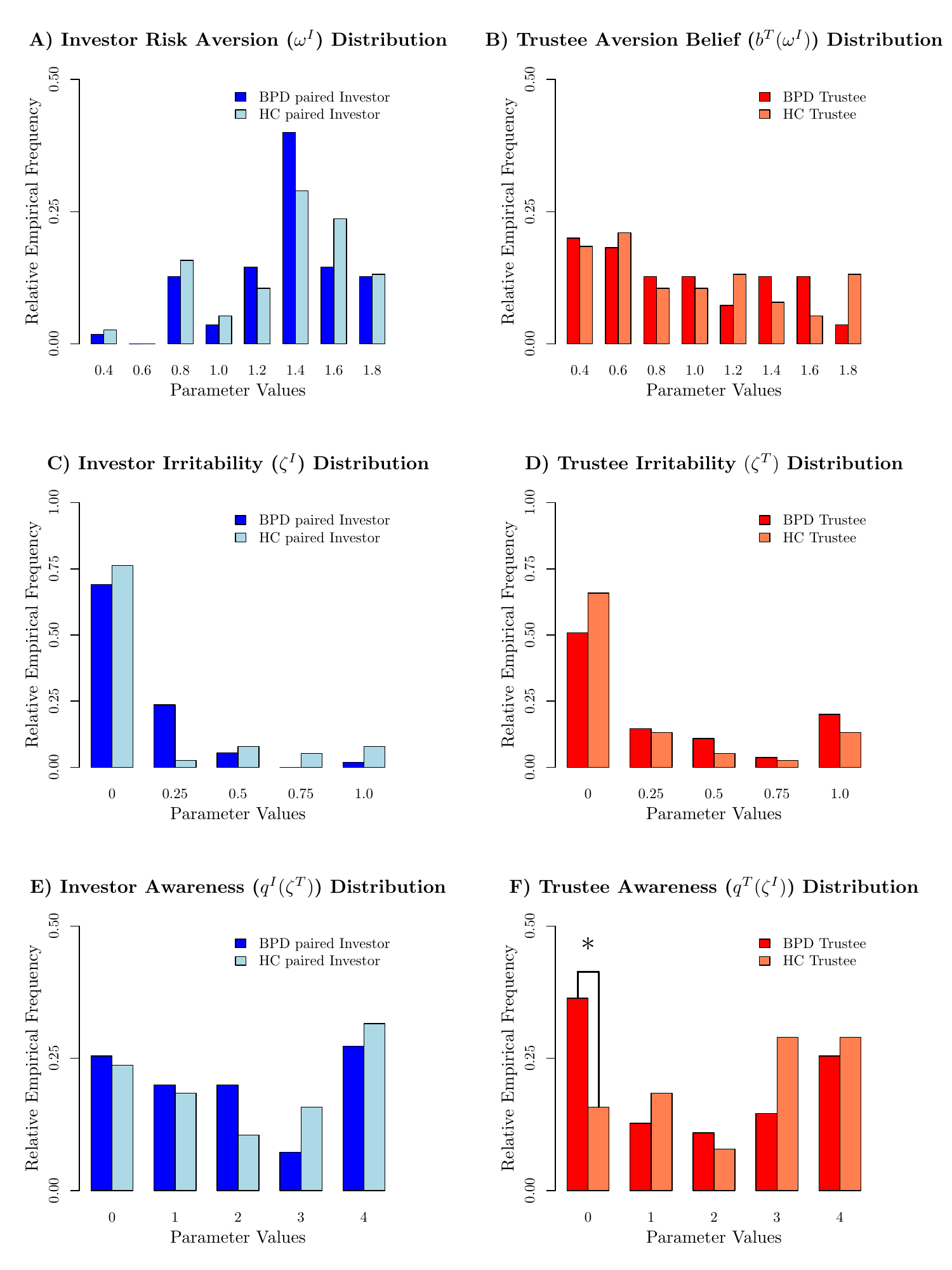} 

	\captionof{figure}{\small{  A) Risk Aversion distribution of investors BPD and HC. B) Risk Aversion distribution of trustees BPD and HC.
C) Irritability distribution of investors BPD and HC. D) Irritability distribution of trustees BPD and HC.
E) Awareness distribution of investors BPD and HC. F) Awareness distribution of trustees BPD and HC. }
	\label{fig:Stat}}
\end{center}

We therefore considered a model-based characterization of the subjects
in which we combined together the two key differences between HC and BPD
trustees in the model: trustees who are either totally guilt-less
($\alpha^T = 0$) or who are irritation unaware ($q^T(\zeta)=0$), or
both. Either of these leads to trustees who will attempt to exploit the
investor, and so create problems in the context of an interaction in
which latter is in charge. We describe these trustees as being
'perilous'.

This group turns out to be present at a significantly ($p=0.0034$,
$\chi^2$-test for equal proportions) higher proportion ($60.0\%$) in the
BPD group of \cite{Brooks2008}, compared with the HC group ($29\%$), the
difference remaining significant ($p<0.05$) even when Bonferroni
correcting for the $10$ ($7$ parameters plus the $2$ proportion tests
and the derived ``perilous group'' hypothesis) comparisons that we
undertook.

Figure \ref{fig:Beh} shows investment and repayment profiles for dyads
in \cite{Brooks2008} including perilous (A) and non-perilous (B)
trustees.  Not only are these interaction profiles evidently different
(confirmed in two-sided t-tests at $p < 0.05$, Bonferroni corrected for
the $10$ time points), but also having adjusted for this by sorting
healthy controls and BPD trustees according to perilousness, there is no
longer a difference between the average investment and return profiles
for BPD versus HC dyads ($p>0.05$ using an uncorrected two-sided
t-test). 

\begin{figure}
\begin{center}
\includegraphics[width=6in, height = 6in]{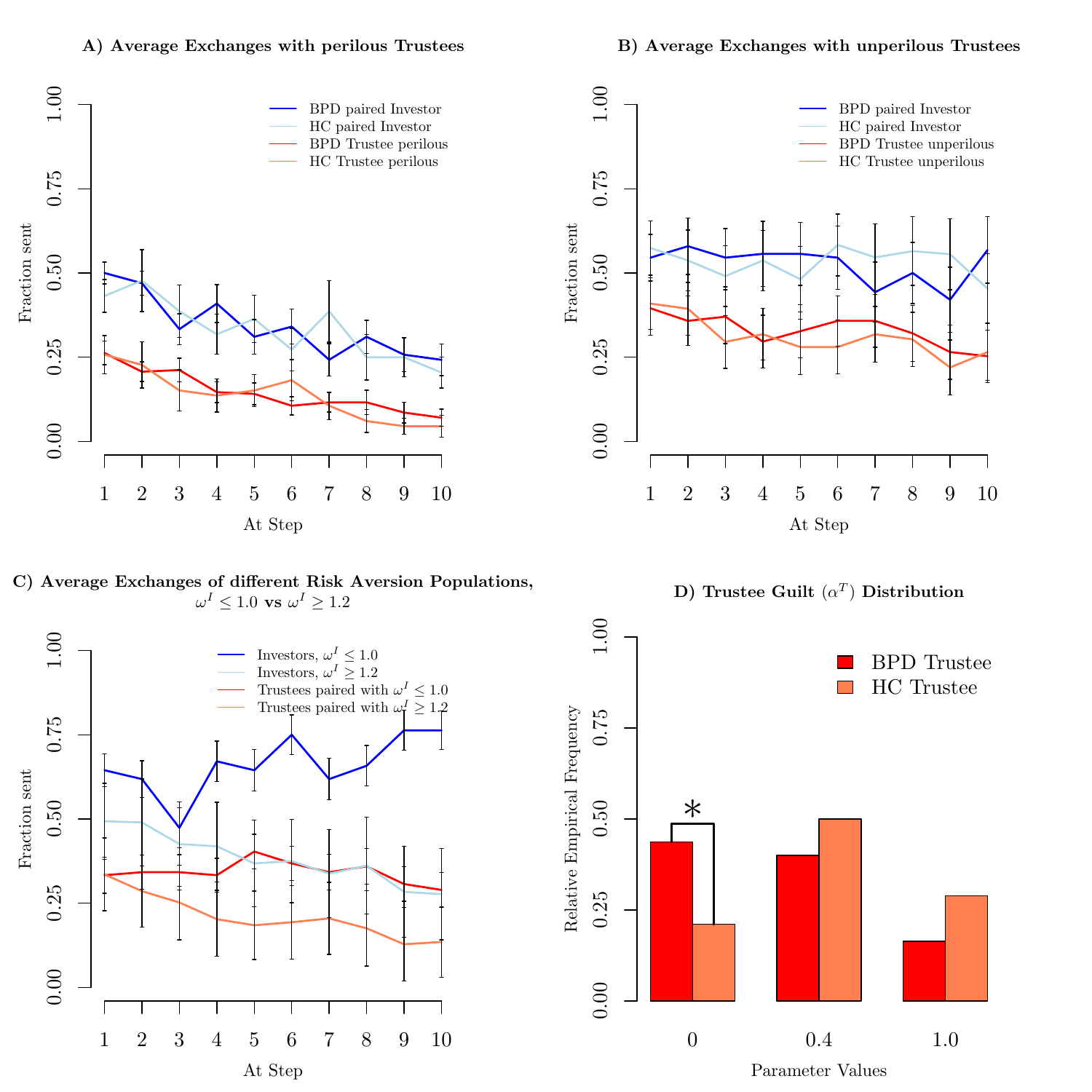}
\end{center}
	\caption{\small{A) Investment and return profile for
            subgroups of the BPD and HC data sets, defined by
            $\zeta^T>0$ or $q^T(\zeta)=0$.  
	B) Investment and return profile for subgroups of the BPD and HC data sets, defined by $\zeta^T=0$ and $q^T(\zeta)>0$. 
	C) Investment and return profile for subgroups defined by $\omega^I \leq 1.0$ (blue, red) or $\omega^I \geq 1.2$ (light blue, coral). 
	 D) Guilt distribution of trustees BPD and HC.
	All errorbars are standard error of the mean. }
	\label{fig:Beh}}
\end{figure}

Figure \ref{fig:Beh}C compares investment and return profiles for
investors with little ($\omega^I \leq 1.0$) or substantial risk aversion
($\omega^I \geq 1.0$).  Splits based on trustee risk aversion profiles $b^T(\omega^I)$ do not appear
significantly different (which is also a testament to the dominant role of the investor) and 
are not shown here. Finally, figure \ref{fig:Beh}D shows the distributions over
the guilt parameters for BPD and HC trustee subjects.

\end{section}

\begin{section}{Discussion}

  Our previous model of the complex collections of choices apparent in
  the multiround trust task did a generally good job at accounting for
  many aspects, and generated prediction errors and other parametric
  regressors that unearthed various key neural processes. However, on
  closer inspection, it failed to characterize aspects of behaviour at
  two disparate timescales: a persistent reluctance of the dominant
  party to submit a portion of their endowment to the potentially fickle
  trustee in the game; and temporary breakdowns in cooperation and
  consequent repair.  We therefore enriched our model in these two
  respects, parameterizing risk aversion (a factor that had previously
  been suggested as potentially corrupting the measurement of trust with
  this task \cite{HouserTrustRisk}), and irritation.

  Despite its formal appeal, the I-POMDP model has not been extensively
  used to characterize game theoretic interactions between players. One
  obvious reason for this is its apparent computational cost. Here we
  showed that it is perfectly possible to perform approximate I-POMDP
  inferencee in a relatively complicated model with two intentional
  dimensions and various other parameters. This augurs well for the
  future, given the importance and richness of social interactions in
  both economic decision-making, and as a psychological biomarker in
  psychiatric conditions.
  
  Our extension of the I-POMDP framework to allow internal state shifts
  (and agents that may be aware of such shifts) adds a crucial layer of
  flexibility to these approaches.  We illustrated this using irritation
  as an elemental emotional process. This captured the rupture and
  repair of cooperation, along with the associated threats of these. In
  the same way that the possibility of punishment or defection maintains
  cooperative behaviour in tasks such as the public goods game, the
  possibility of rupture encourages healthy participants to be
  beneficent. We hope that similar mechanisms will also be useful to
  describe strategic interactions in other tasks. We will also use it to
  guide the analysis of functional brain imaging data.

  This model of irritation departs from conventional models of
  intentional inference in one important way. In repeated social
  exchange tasks, it is conventional to model one's partner's
  \emph{preferences}, which, in Bayes-Nash terms, concerns properties of
  their \emph{utility functions}.  Indeed, this is exactly how earlier
  studies on the multi round trust game framed the social exchange
  \cite{Debbs2008, Misha2010,Ting2012,Hula2015}. Here, however, we
  considered simultaneous intentional inference about both a
  \emph{utility} and a \emph{policy} (as in \cite{Wunder}) that the
  player would adopt (indeed, a policy that it would be hard to justify
  in pure utility terms, given the costs of breaking cooperation). We
  include the possibility of one agent's actions changing the
  \emph{intentional state} of another agent, thus extending beyond the
  non manipulability assumption in \cite{iPOMDP} and providing a
  tracktable time series of irritation/state shifts (see figure
  \ref{fig:Brooks}AB). This non-stationarity could be accommodated
  within the parameterized I-POMPDs of \cite{Wunder}, using a
  specially-fashioned extension of the intentional state space.
  
  A richer palette of such internal state-shifting default
  behaviors might also prove important in other tasks. Note, though,
  that it is not yet clear that a suitable notion of equilibrium can be
  defined (for instance, as the theory of mind level of the players
  tends to infinity). The combination of Kuhn's theorem \cite{Kuhn}
  (since our players have perfect recall) and Harsanyi's treatment of
  mixed strategies \cite{HarsMixed} might be a starting point
  for such a treatment.

  Our approach to irritability was chosen for its simplicity within the
  existing model.  Further work on a more substantial body of human data
  will be necessary to fine-tune the dynamics of irritation in social
  exchange. One first step might be to use the model as part of an
  optimal experimental design framework to realize a computer-based
  opponent that could extract the most out of each available choice.  At
  present, the relatively small number of actions in our version of the
  trust task, together with the possibility that the human partners fail
  to irritate each other even when they are irritable, leaves little
  room for further sophistication. Given a better understanding of
  irritation in the model, it would then be possible to refine the
  concept itself.
  
  The ultimate model has the uncomfortable characteristic of employing
  $7$ parameters to account for the $10$ choices of each subject.
  However, the parameters interact in complex ways in the model, which
  is why they can generally be reliably inferred, as apparent in the
  confusion matrices in the Supplementary materials.
 
  Finally, the model provides a generative approach to the way that
  patients with Borderline Personality Disorder play in the multi round
  trustgame, as reported in \cite{Brooks2008}. This approach yields a
  particular type of trustee, who are perilous for the interaction;
this type was overrepresented in the BPD sample. 
After taking proper account of this subtype, we found equal average
behaviours in BPDs and HCs. Thus this subgroup (which is also present in
the HC group, albeit to a lesser extent) could help separate out a
clinical phenotype that is separate from those sufferers of BPD who are
less susceptible to the breakdown of trust. Such a separation might
yield clearer clinical and neurological characterisations. It would be
most interesting to look for, and analyze the clinical correlates of,
types analagous to perilousness for players who are in control of
interactions, like the investors here.

\end{section}

\begin{section}{Acknowledgements}
  We thank Terry Lohrenz and Michael Moutoussis for
  helpful
  discussions and Tobias Hauser and Michael Moutoussis for comments on the manuscript.
  Special thanks go to the IT support staff at the Wellcome Trust
  Center for Neuroimaging and Virgina Tech Carilion Research
  Institute. The authors gratefully acknowledge funding by the
  Wellcome Trust (Read Montague) under a Principal Research
  Fellowship, the Kane Foundation (Read Montague) and the Gatsby
  Charitable Foundation (Peter Dayan). Andreas Hula is supported by
  the Principal Research Fellowship of Professor Read Montague.
\end{section}
\newpage
\bibliography{NeurBibAlt}
\bibliographystyle{apalike}{}


\renewcommand{\thefigure}{S\arabic{figure}}
\renewcommand\thesection{S}   
\setcounter{figure}{0}  
\begin{section}{Supplementary Material}

\subsection{The Multi Round Trustgame Model}\label{OldModel}

An I-POMDP generative model for the trust task was proposed in
\cite{Debbs2008,Ting2012} and substantially refined in \cite{Hula2015}.
The
proximal cause of behaviour is a set of reward expectations $Q (a , h)$
for taking a given action $a$ after having experienced a history of
events $h$ in the game. Here the agent was supposed to learn about the
other agent from this history $h$, following Bayes rule from a given
initial belief system. These are assumed to generate choices using a
softmax rule (something that is known to all parties)
\cite{Norm, 2step, nogo,quantal} 
\begin{equation}\label{eq:softmax}
\pi(a,h)=\mathbb P  [ a | h] =  \frac {e^{ \beta Q (a , h)}}
{\sum_{b\in\mathcal A} e^{\beta Q(b , h)} } 
\end{equation}
where $\beta>0$ is called the inverse temperature parameter
  and controls how diffuse are the probabilities. The policy
\begin{equation}\label{eq:argmax}
\pi(a, h) = \left\{
\begin{array}{ll}
1 & \textrm{ if }  Q (a , h) = \max{  \{   Q (b , h) | b \in
  \mathcal A \}} \quad \textrm{(assuming this is unique)} \\[2pt]
0 & \textrm{otherwise}
\end{array} \right.
\end{equation}
can be obtained as a limiting case for $\beta \rightarrow \infty$.

The generative model parameterized the interaction using three
non-intentional parameters: inverse temperature, theory of mind level
and planning horizon; along with guilt as an intentional parameter
(which is subject to intentional reasoning). A limited range of values
was assumed for each parameter, covering the observed behaviour.

The inverse temperature $\beta$ in the softmax
decision making mechanism
could take values $\{ 1, \frac 1 2, \frac 1 3, \frac 1 4 \}$.

The theory of mind (ToM) level $k$ (see \cite{Costa}) encodes how many
recursive steps an agent uses in their intentional model of the partner
i.e. a level $0$ only learns about the partner, a level $1$ knows that
the partner is learning about them and so forth. Participants know 
their own levels ($k^I, k^T$) and
assume that their partners play at one level lower (so the investor
thinks the trustee has level $k^I-1$; the trustee thinks the investor
has level $k^T-1$). 

Recursion starts with a minimal model, dubbed $k=-1$, assumed to
play a static strategy based on immediate preference (see
\cite{Hula2015}). For consistency, this implies that the level $-1$
models do not themselves hold intentional models of their partners
anymore, but act as in a non-intentional environment. An alternative
approach to levels of thinking would be akin to the cognitive
hierarchy, suggested in \cite{Camerer}, allowing higher level 
agents to maintain beliefs about multiple possible partner levels below them, 
instead of fixing the partner model at one less than their own level. 

Based on the observation \cite{Hula2015} that only even investor levels
and odd trustee levels produce distinct behaviours, we consider $k^I \in
\{ 0, 2, 4, 6, 8 \}$ for investors and $k^T\in\{0 , 1 , 3, 5, 7\}$ for
trustees (the original model was restricted to the first two in each
set). 

The planning horizon $P$ is the number of steps of future interaction
taken into account in determining $Q(a,h)$. We consider $P \in
{1,2,3,4}$ (the original model considered $\{0, 2, 7 \}$; however,
behaviour for $P=0$ is very similar to $k=0; P=1$, since $P=0$ implies
$k=0$ in this type of model-interaction, and $4$ lowers memory costs
considerably).  A discussion of why we consider this level of planning 
sufficient can be found below.

The intentional parameters $\alpha^I, \alpha^T\in\{0,0.4,1\}$ quantify
guilt (see \cite{Fehr}), and change subjects' utility functions from
those described above for the investor to
\[
r^I (a^I, a^T,
\alpha^I) =\chi^I(a^I,a^T) -\alpha^I \max \{ \chi^I(a^I,a^T) -\chi^T
(a^I, a^T),0\},
\]
 and for the trustee to
\[
r^T(a^I, a^T, \alpha^T) =
\chi^T(a^I,a^T) -\alpha^T \max \{ \chi^T(a^I,a^T)-\chi^I(a^I,a^T) ,
0\}.
\]
High guilt $\alpha=1$ means that every point of advantageous inequality
in payoffs diminishes the utility of the outcome by $1$ i.e. there is no
felt benefit from having a larger payoff than the partner.  A low guilt
of $0$ means that only raw outcome maximization is relevant to the
agent, while $0.4$ is a more measured, but mostly self-interested agent.

Agents know their own guilt; but
adopt a multinomial distribution on the possible guilt values of their
partner, with a Dirichlet prior on probabilities of the multinomial
distribution.  Thus the initial belief state ($\mathcal B_0$) is a
symmetric Dirichlet-Multinomial distribution,
\[
\mathcal B_0 \sim DirMult(d_0),
\qquad d_0 = (1,1,1).
\]
The level $k=-1$ agent assumes equal probability for all partner guilt
states and neither learns nor plans for the partner's decisions, only
using immediate expected utilities.  However, the more complex agents
learn and make recursive inferences about their partners. 
To be consistent with \cite{Ting2012}, the posterior
distribution is approximated as a Dirichlet-Multinomial distribution
with the parameters of the Dirichlet prior being updated to
\[
d_{t+1}^{i}=d_t^{i} +\mathbb
P[o_{t+1}=\textrm{observed action}| \alpha^{\textrm{partner}}=\alpha_i].
\]

In sum, a baseline for any I-POMDP model is random choice, defined by
a probability of $20$ percent for each possible action. In likelihood
terms, over the $10$ interactions in the multi round trust game this
would correspond to a negative loglikelihood (NLL) of $16.1$ nats.
 The parameters $\theta_M$ characterizing a subjects'
play under the given model $M$ were determined by minimizing the NLL
under the generative model $M$.

\subsection{Theory of Mind Limitation}\label{OldToM}
  In figure ~\ref{fig:ToM8}, we display average interactions of
  investor and trustees, with basic parameters: $\alpha^I=\alpha^T =
  0.4$, $P^I=P^T = 4$, $\beta^I = \beta^T = \frac{1}{2}$ and various
  theory of mind levels. Investors and Trustees are always one level
  apart. In the left column the investor has the higher level and in
  the right column the trustee has the higher level. Since their guilt
  is $0.4$ the trustee will try to exploit the investor. In the left
  column the investor is of a higher level and therefore sees through
  the trustee's deception and lowers the investment. In the right
  column the trustee is of higher level and successfully exploits the
  investors. From the level $4$ investor (figure ~\ref{fig:ToM8}C)) the
  behaviours do not appear to be qualitatively different anymore,
  except for minor differences in rates of increase of decline, which can 
  be accounted for by the new parameter of risk aversion and 
  inverse temperature (see the main 
  text).
  Therefore we restricted our data fit to a maximum ToM level of $4$.

\begin{center}
\includegraphics[width=5in, height = 8in]{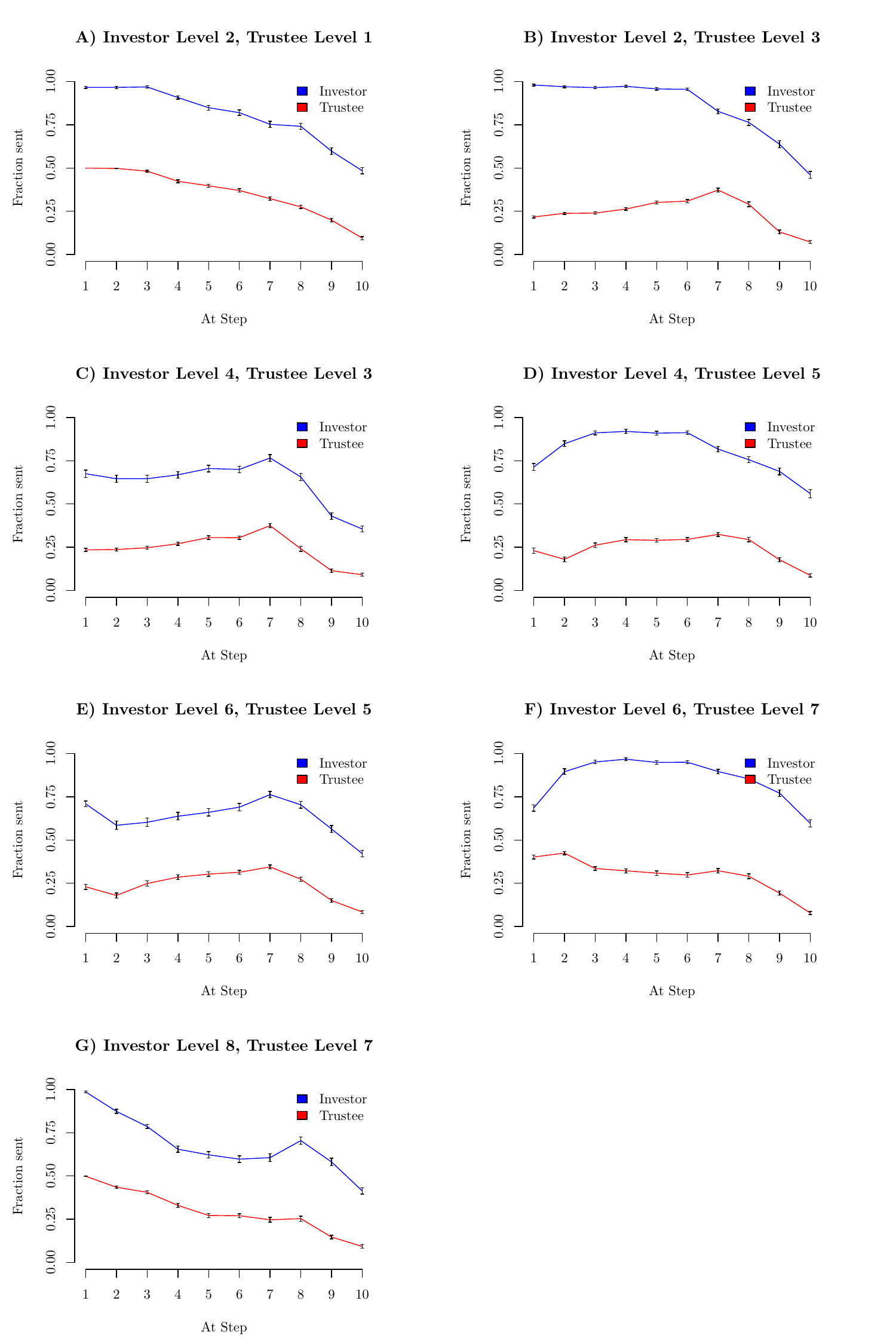}
	\captionof{figure}{\small{Averages of $200$ sample paths, displaying interactions between 
	investors and trustees one level apart, with parameters $\alpha^I=\alpha^T = 0.4$, $P^I=P^T = 4$, $\beta^I = \beta^T = \frac 1 3$ . left column: Investor Level is higher. Right column: Trustee 
	Level is higher. All errorbars are standard error of the mean.}
	\label{fig:ToM8}}
\end{center}
\subsection{Planning}\label{OldPlan}

In this section, we confirm that we can recover the paradigmatic
behaviours of \cite{Hula2015} with planning $4$ as well as we could with
planning $7$ in the earlier work.  The essential behaviours can be seen
in figure ~\ref{fig:Plan}A-D. 
Figure ~\ref{fig:Plan}A is based on a level $2$ investor and a level $1$
trustee, with $\alpha^I=\alpha^T = 0.4$ and $\beta^I = \beta^T = \frac 1
3$ and $P^I=P^T = 4$: The trustee tries to build up trust and then
exploit the investor. The investor being level $2$ is not deceived by
the trustee and reduces their investment, as the trustee defects, thus
\emph{preempting exploitation}.  Conversely in figure ~\ref{fig:Plan}B,
the investor is level $0$ and $\beta^I = \beta^T = \frac 1 3$, 
with the other parameters being the same. In this case,
the trusee successfully builds trust in the first few exchanges and
later on exploits the investor, who keeps giving till the very last
exchanges, believing the trustee to trustworthy. In figure
~\ref{fig:Plan}C the ``impulsive'' behaviour of \cite{Hula2015} is
repoduced: It shows the average exchanges of a level $0$ investor and a
level $1$ trustee, with $\alpha^I=\alpha^T = 0.4$ and $\beta^I = \beta^T
= \frac 1 3$ and $P^I=P^T = 2$. The planning horizon leads the trustee
to exploit the investor too early, therefore earning much less than in
the case of figure ~\ref{fig:Plan}B, despite being also of higher level
than the investor and having a matched planning horizon. Finally, figure
~\ref{fig:Plan}D demonstrates the importance of the planning horizon by
showing average exchanges of a level $2$ investor and a level $1$
trustee, with $\alpha^I=\alpha^T = 0.4$ and $\beta^I = \beta^T = \frac 1
3$ and $P^I=2$ but $P^T = 4$: Although the investor is of a higher
level, the longer planning/more consistent trustee successfully deceives
them, with the light drop off in investments in the end being more due
to horizon effects than to the investor looking through the trustee's
deception.

We do not present the fully cooperative (guilt $1$) case as well as the fully greedy case (both partners guilt $0$) since those are essentially unrelated to depth of planning.
\newpage
\begin{center}
\includegraphics[width=6in, height = 6in]{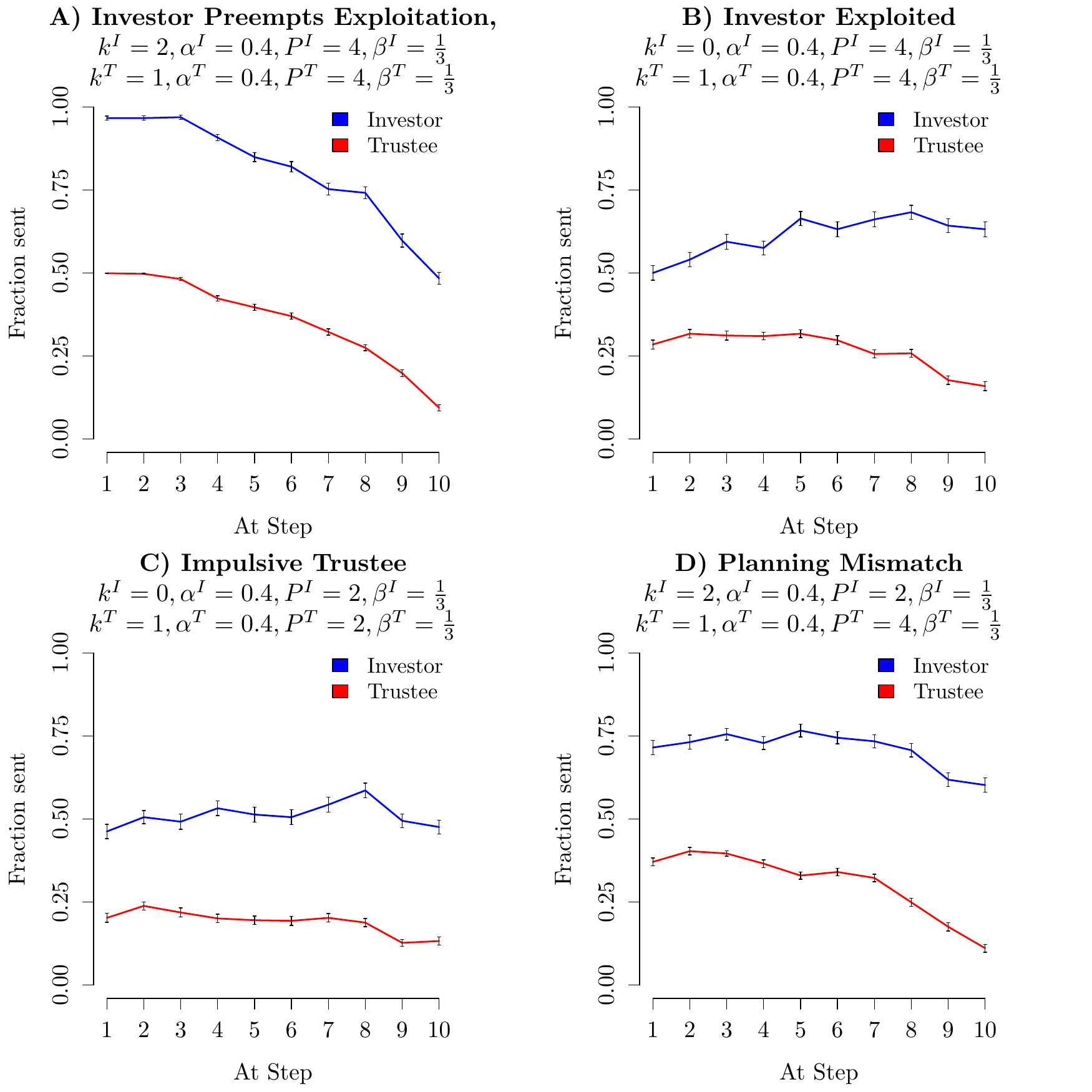}
	\captionof{figure}{\small{Averages of $200$ sample paths. A) Investor reacts to exploitation by 
	 trustee . B) Investor is exploited by 
	 trustee . C) Trustee is too inconsistent 
	 to exploit Investor D) Trustee exploits 
	 higher level shorter planning Investor. All errorbars are standard error of the mean.}
	\label{fig:Plan}}
\end{center}
In the face of these reproductions, we can conclude that all paradigmatic
behaviours of \cite{Hula2015} can just as well be produced with
planning $4$ as with planning $7$.

\subsection{Risk Aversion}\label{RiskDetail}

The effect of risk
aversion in shifting investment levels can be seen in
figures~\ref{fig:Aversion}A;B. 
These depict the average investment (A)
and repayment (B) trajectories over $200$ simulated exchanges in which a
trustee with $b^T(\omega^I)=1$ (i.e., who believes the investor to have
$\omega^I=1$) interacts with investors of varying actual $\omega^I$
values (other parameters are given in the caption).  Cooperative trustee
actions early in the game can make the investor overcome moderate levels
of risk aversion (the curve for $\omega^I=1.2$ merges with the curve for
$\omega^I=0.8$ in the early trials). Higher risk aversion levels
($\omega^I=1.4$) delay the positive effects of cooperation, such that it
increases from a low initial level until step $6$, but then drops
abruptly due to horizon effects and trustee defection.  For the highest
risk aversion levels $(\omega^I=1.6;1.8)$ in figure
~\ref{fig:Aversion}A, inference about other parameters may be hampered.
That is, if investments stay low throughout, risk aversion might become
nearly the only parameter that can be inferred with certainty. This
implies a constraint on any further statistical treatment of behavioural
data or derived quantities, such as model-based fMRI analysis.

Figure ~\ref{fig:Aversion}C depicts the effect of the risk aversion
belief on the trustee, with the investor now being fixed at $\omega^I =
1.0$ (and not shown). For $b^T(\omega^I) < 1.0$, it can be seen that
trustees make early attempts to exploit investors they think are not
risk averse, since they assume the investor is still more likely to
invest in them than to defect. However, they then defect rather quickly.
As the trustee's belief approaches the utilitarian setting
$b^T(\omega^I) = 1.0$, she becomes more cooperative and for a longer
time period into the game, since this is necessary to keep the
investments of the utilitarian investor going.  Given a high setting of
$b^T(\omega^I) = 1.4$ the trustee returns little, as they consider the
likelihood of the investor to keep investing in them to be low and thus
see no gain from building up cooperation. 

\begin{figure}
\begin{center}
\includegraphics[width=5in, height = 7in]{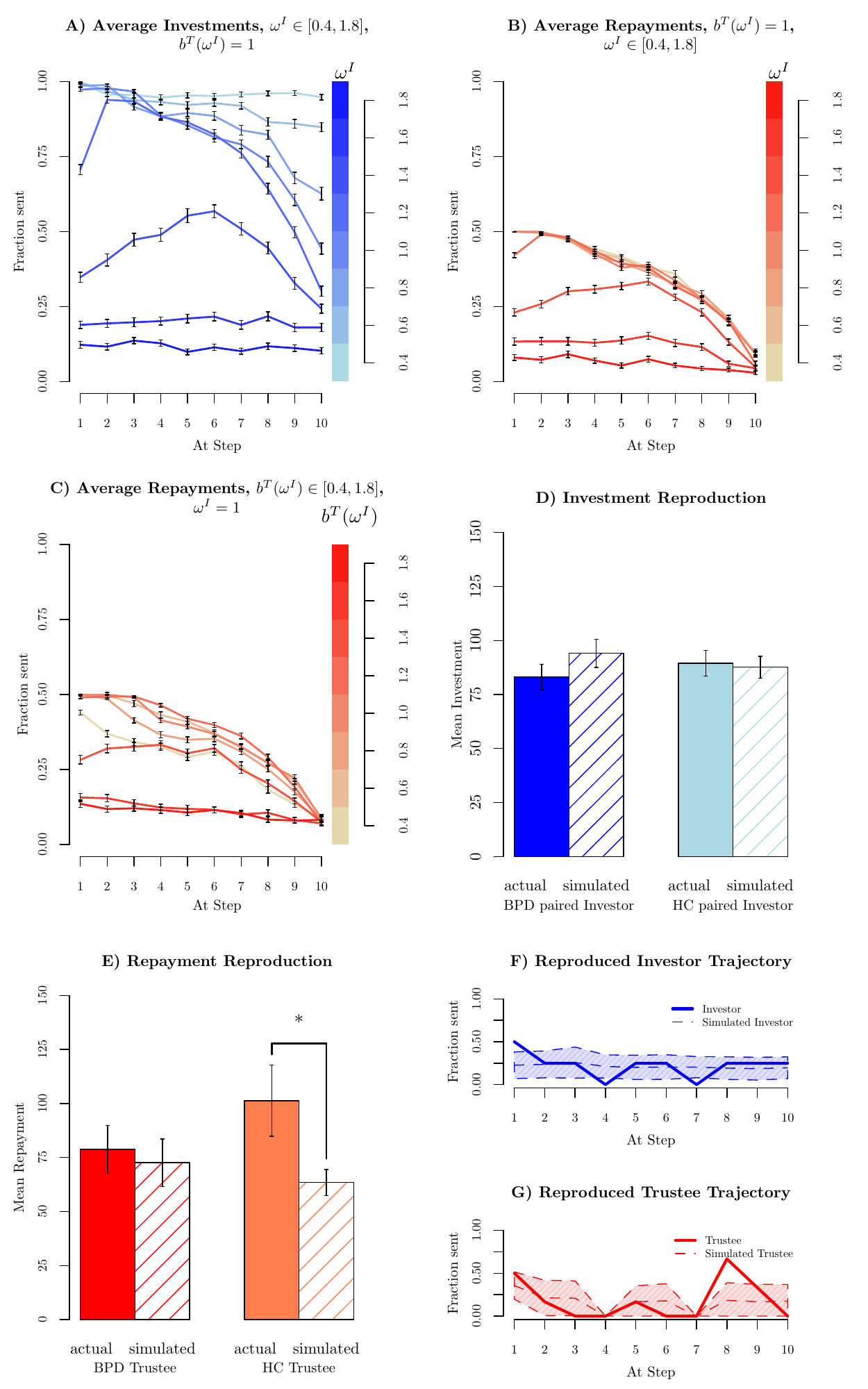} 

	\captionof{figure}{\small{A) Investment profiles for different
            settings of investor risk aversion. All errorbars are
            standard errors of the mean. Here, the investor and trustee
            have guilt $\alpha= 0.4$, inverse temperature $\beta=\frac 1
            2$, and planning $P=4$; the investor (respectively trustee)
            has ToM level $k^I=2$ ($k^T=1$). B) Trustee repayment
            profiles for the interactions depicted in A.  C) Repayment
            profiles for different settings of trustee risk aversion
            belief. All errorbars are standard errors of the mean. Here,
            the investor and trustee have guilt $\alpha= 0.4$, inverse
            temperature $\beta=\frac 1 2$, and planning $P=4$; the
            investor (respectively trustee) has ToM level $k^I=2$
            ($k^T=1$).  D) Average Investment profiles regenerated from
            estimated parameters. All errorbars are standard error of
            the mean.  E) Average Repayment profiles regenerated from
            estimated parameters. All errorbars are standard error of
            the mean.  An asterisk denotes a significant difference
            ($p<0.05$, two sided t-test) in means between the original
            data and the generated exchanges.  F) Sample investor
            trajectory vs average of $200$ generated exchanges using the
            model augmented by risk aversion. Shaded areas are estimated
            standard deviation.  	 G) Sample trustee trajectory vs
            average of $200$ generated exchanges using the model
            augmented by risk aversion. Shaded areas are estimated
            standard deviation.  }
	\label{fig:Aversion}}
\end{center}
\end{figure}

As figures~\ref{fig:Aversion}D;E demonstrate, the investment can on
average be well reproduced after including risk aversion in the model,
although there remains a significant under-reciprocation on the part of
the generated HC trustee, compared with the real exchange data.
Further, figure~\ref{fig:Aversion}FG demonstrates that despite an
improvement in fit (NLL for the investor goes from $14.04$ to $7.47$
with $\omega^I=1.4$; for trustee the NLL goes from $11.93$ to $11.53$
with $b^T(\omega^I)=0.6$), this model remains incapable of capturing the
transient rupture and - by extension - repair that we examined in
figure~\ref{fig:Brooks}D.  Again, the modelled investor decreased their
investment to $0$ on only $23$\% of the sample runs on trials $4$ and
$7$, compared with the collapse in the actual investment.

\subsection{Quantitative Illustration of Irritability}\label{IrrQuant}
Figure ~\ref{fig:Repair} shows the effect of irritability and
irritability inference on average behaviours over $200$ simulated
exchanges. 
As is also evident in figure~\ref{fig:Brooks}, these
averages blur the precise times at which the ruptures happen, but show
the consequences in terms of net cooperation.  In both cases (A-C;
D-F), the trustee is more sophisticated than the investor ($k^I=0;
k^T=1$) , the investor is either irritable or non-irritable, but
unaware; the trustee is nonirritable (other parameters are listed in
the caption).  The difference between the figures is that the trustee
is aware in figure~\ref{fig:Repair}A-C, but unaware in figure
~\ref{fig:Repair}D-F.

Figure~\ref{fig:Repair}A show average investment and returns when the
trustee is aware for an irritable (dark) and non-irritable (light)
investor. The trustee's awareness enables her to keep the investment at
almost the same level in both cases.  This arises from the excess return
that she provides. Thus, as in figure~\ref{fig:Brooks}F, trustees who
realize that their investors are irritable will delay exploitation to the
late rounds of the game. For the case of the irritable investor,
figure~\ref{fig:Repair}B shows the average evolution of the inference
about partner irritability.  The trustee becomes aware that their
partner is likely irritable after the first retaliation (as in the
particular case in figure~\ref{fig:Brooks}F).  The average internal state
of the same investor and trustee can be seen in figure
\ref{fig:Repair}C.  Overall the irritation weight of the investor is
kept low by an aware trustee, who can repair the interaction if needed.
The value of irritability between investor and trustee is not symmetric
in this example, so that the trustee may reliably repair or fail to
repair without being themselves subjected to the effects of irritation.
We note that the investor is driven up to near equally high investments
in the non-irritable case in figure ~\ref{fig:Repair}A compared to that
in the irritable case of figure~\ref{fig:Repair}A, despite the trustee
actually returning less.  This is because the level $k^T=1,
q^T(\zeta^I)=0$ trustee knows exactly what actions they need to take in
order to confuse the inference of the level $k^I=0$ investor (i.e.  what
responses the investor will consider unlikely).  By contrast, the level
$k^T=1, q^T(\zeta^I)=1$ trustee in figure~\ref{fig:Repair}A accounts for
potential irritability right away and thus has to ``play along'' with
the investor $k^I=0$'s expectations and is less effective in tricking
their inference.

\begin{center}
\includegraphics[width=6in, height = 4in]{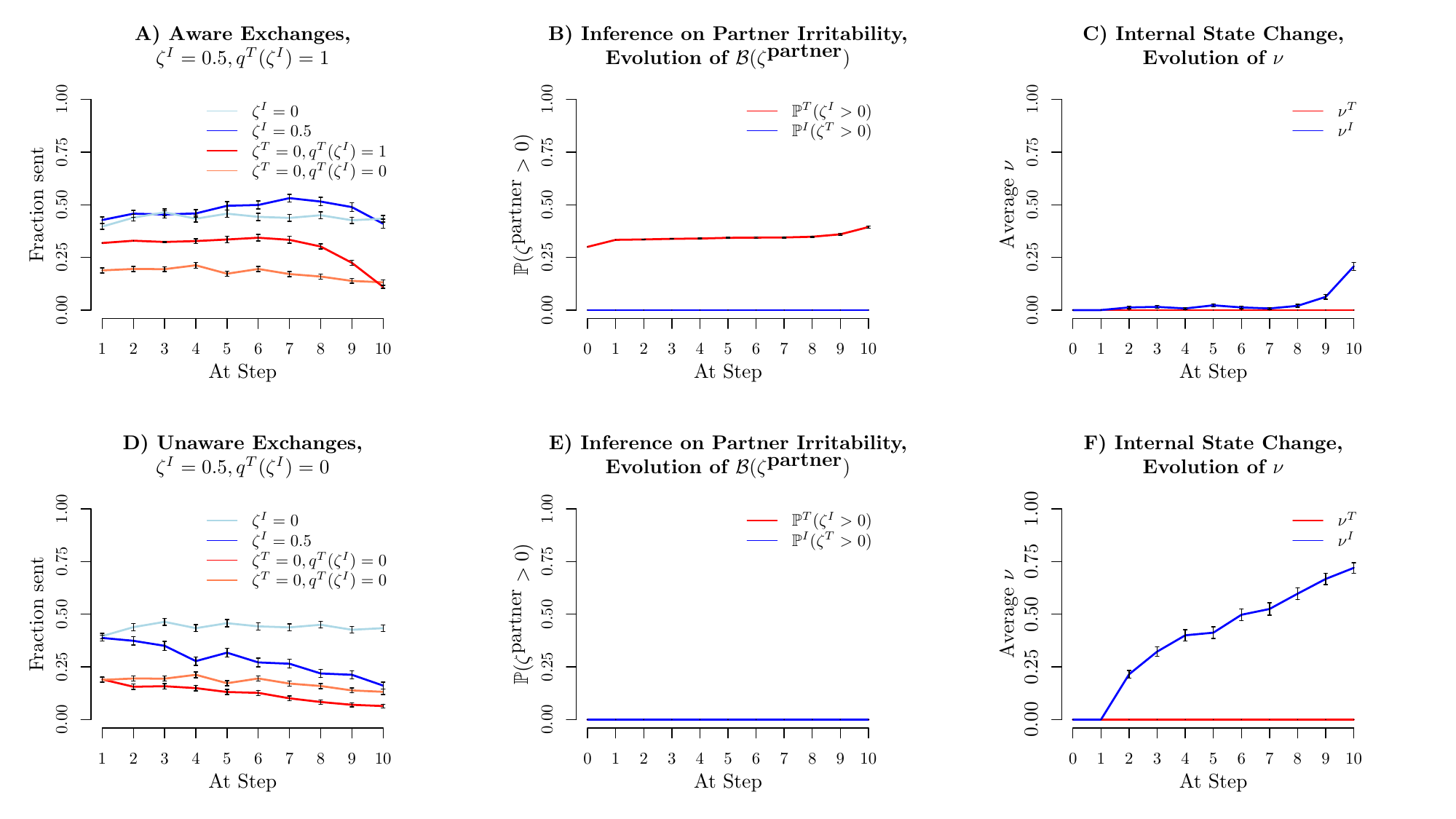}  
\captionof{figure}{\small \small{A-C) Simulated repair interaction with
    an irritability aware trustee ($q^T(\zeta^I)= 1)$ and unaware investors who
    are irritable (dark lines; $\zeta^I = 0.5$) or non-irritable (light
    lines; $\zeta^I=0$). A) Average investment profiles in the two
    cases. B) Average evolution of the irritability beliefs of both
    partners. The trustee learns correctly that the irritable investor
    is irritable. C) Average evolution of the irritation weight
    $v_{\iota}$ for both partners. The awareness keeps the irritation
    weight low and the trustee can repair the interaction if needed.
    D-F) Simulated breaking interaction with the same investors, but an
    unaware trustee ($q^T(\zeta^I)=0$). The plots show the same quantities as
    in A-C. All errorbars are standard errors of the mean
    over $200$ simulations. Investor/trustee parameters are $k=0/1;
    \alpha=0.4; \omega=1.4; P=4; \beta=\frac 1 2$ and the trustee had the
    fixed belief that the investor's risk aversion was $\omega=1.4$.  }
	\label{fig:Repair}}
\end{center}

By contrast, figure~\ref{fig:Repair}D shows that if the trustee is
unaware, then there can be ruptures of cooperation that become
apparent even at the group average level.  The evolution of the
beliefs about irritation is nugatory, as can be seen in
figure~\ref{fig:Repair}E.  Figure~\ref{fig:Repair}F shows how the
irritation weight reaches high values quickly, only occasionally being
reduced by chance repair. 

\subsection{Parameter Recoverability}\label{ModelInversion}

The ultimate model is rather complicated. This raises the concern that
the same behaviour might result from radically different settings of
the parameters, implying that we would not be able to draw stable or
meaningful conclusions from fitting behaviour. Indeed, we have already
observed that certain settings will make it impossible to make
inferences about some parameters -- thus, playing with a highly risk
averse investor will give no opportunity for a trustee to express
her individual characteristics.

To examine this, we assessed parameter recoverability. That is, we
used the parameters obtained from ML estimation on the participants
(note that not all values were represented in the population --
$\omega^I=0.6$ is absent, for instance), 
generated new data \emph{ab initio} from the model, fitted the new
data, and quantified any discrepancies between the original and
recovered parameters.
Figures~\ref{fig:InvestorConf}~and~\ref{fig:TrusteeConf} show the
probability of recovering either the actual or a neighboring parameter
value for investor and trustee respectively. 

It is apparent that the model has some significant purchase on all the
parameters. However, some parameters are much harder to estimate than
others. 
There are perhaps four most egregious forms of confusion.  First,
irritable subjects can be inferred as being non-irritable
(figures~\ref{fig:InvestorConf}F;~\ref{fig:TrusteeConf}F).  This occurs
if the remaining randomness of the interaction in the model is such that
the investor's irritation is not excited. Indeed, the task was not
designed with irritation in mind, and so players are not forced or
encouraged to irritate each other.

Second, the investor's awareness is not very reliably recovered
(figure~\ref{fig:InvestorConf}E).  It is slightly better
recovered for the trustee (figure~\ref{fig:TrusteeConf}E), who faces a
more stringent challenge to keep the investor trusting them.

Thirdly, the inverse temperatures $\beta^I$ of generated investor trajectories tend to 
be overestimated (figure~\ref{fig:InvestorConf}B). This is not surprising since the preferred actions will be the same 
for several temperature settings, thus requiring several ``unlikely'' actions to 
identify a lower $\beta^I$ in investors.

Finally, there is a tendency to missestimate the risk aversion belief of
the trustee, for high settings of $b^T(\omega^I)$, as the parameter
apparently has less influence on trustee choice, than it does for the
investor, who is in control of the interaction (see
figure~\ref{fig:TrusteeConf}G). This effect may be driven by the fact
that the trustees estimated to have such high beliefs were also
estimated to be considerably more irritable and thus the ensuing breaks
might have confounded the inference about the risk aversion
belief or vice-versa per chance lower repayments could have been 
interpreted as irritability. Globally, irritability is under rather than overestimated, 
however we can not rule out a particular interaction for high $b^T(\omega^I)$ subjects.

Also, many parameter values which are less reliably recovered are also the less common 
values in the estimated data to begin with, thus making their recovery subject to a higher volatility 
by means of lower numbers.

\begin{center}
\includegraphics[width=5.5in, height = 7.5in]{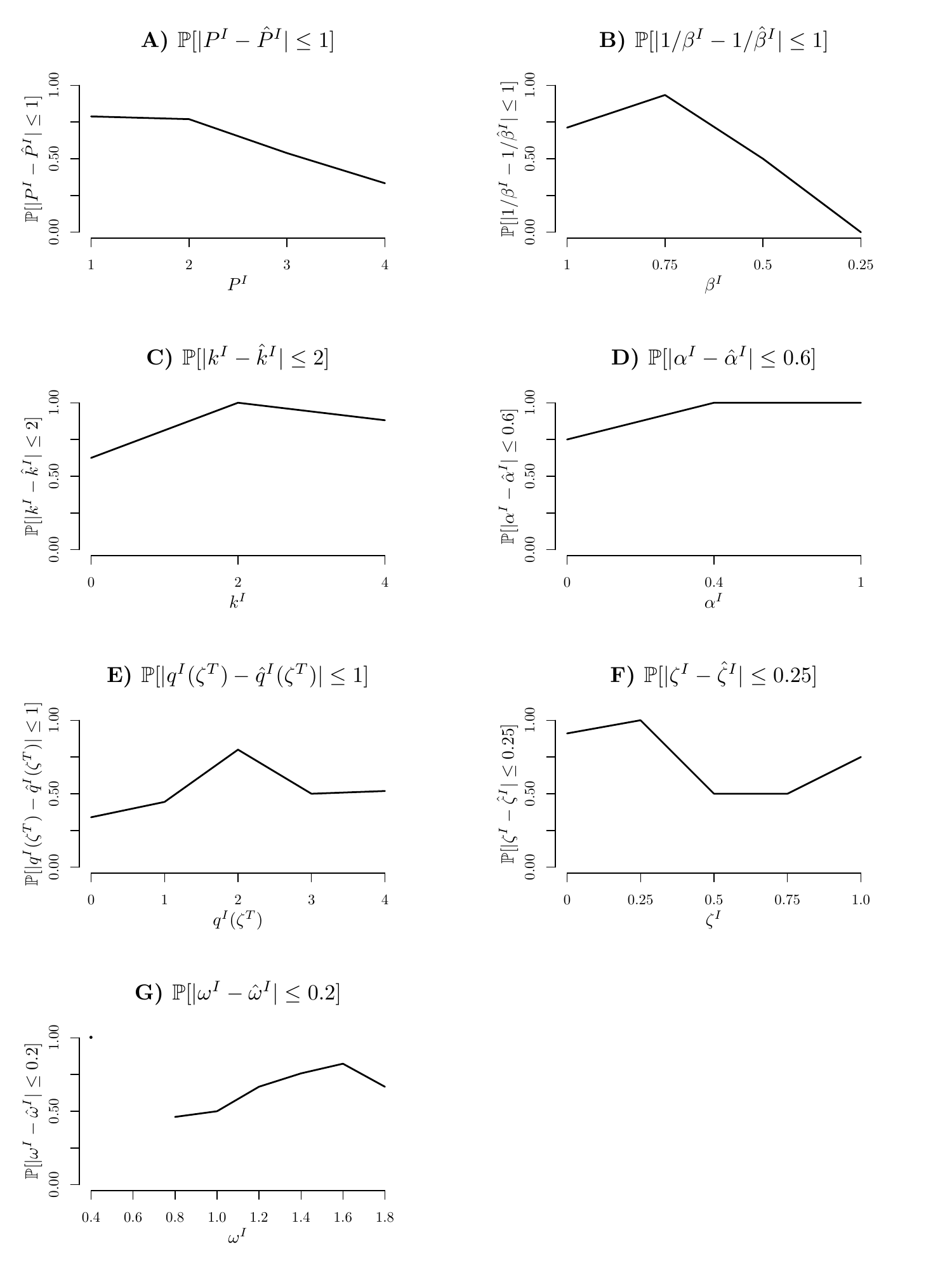} 

	\captionof{figure}{\small{ For the investor: Probability of Parameter recovery or 
	recovery of a neighbouring parameter value from generated exchanges 
	using real subject parameters. A) Probability to recover 
	the planning value $P^I$ or a neighbouring one. B) Probability of recovering 
	$\beta^I$ or a neighbouring value. C) Probability of recovering ToM $k$ or 
	a neighbouring value. D) Probability of recovering Guilt $\alpha^I$ or 
	a neighbouring value. E) Probability of revovering $q^I(\zeta^T)$ or a 
	neighbouring value. F) Probability of recovering $\zeta^I$ or a neighbouring value.
	G) Probability of recovering risk aversion $\omega^I$ or a neighbouring value. The 
	value $\omega^I = 0.6$ did not occur in the original data set.
	}
	\label{fig:InvestorConf}}
\end{center}

 \begin{center}
\includegraphics[width=5.5in, height = 7.5in]{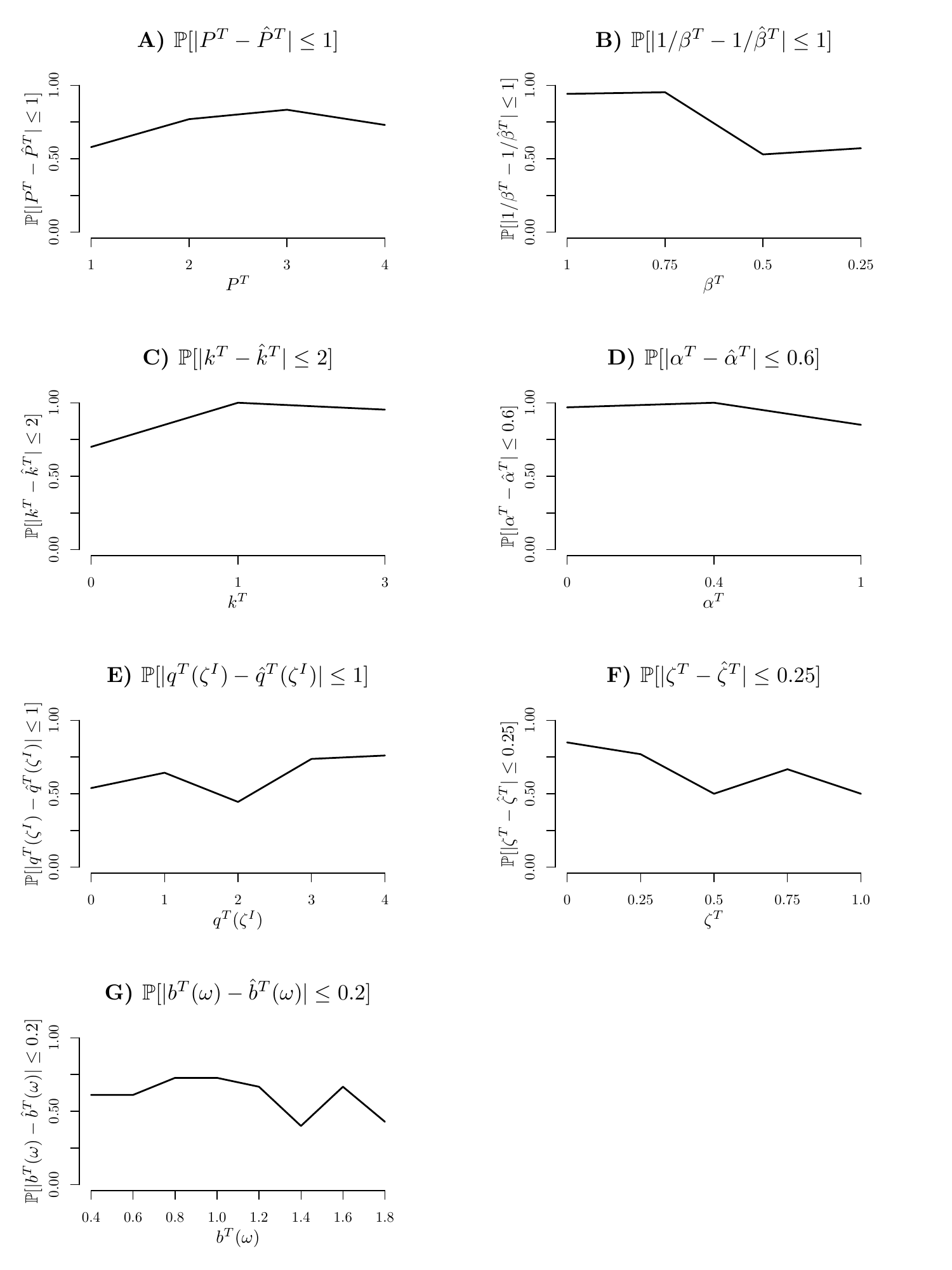} 

	\captionof{figure}{\small{  For the trustee: Probability of Parameter recovery or 
	recovery of a neighbouring parameter value from generated exchanges 
	using real subject parameters. A) Probability to recover 
	the planning value $P^T$ or a neighbouring one. B) Probability of recovering 
	$\beta^T$ or a neighbouring value. C) Probability of recovering ToM $k$ or 
	a neighbouring value. D) Probability of recovering Guilt $\alpha^T$ or 
	a neighbouring value. E) Probability of revovering $q^T(\zeta^I)$ or a 
	neighbouring value. F) Probability of recovering $\zeta^T$ or a neighbouring value.
	G) Probability of recovering risk aversion $b^T(\omega^I)$ or a neighbouring value. 
	}
	\label{fig:TrusteeConf}}
\end{center}

\subsection{Algorithmic Representation}\label{Algo}
We outline the algorithm through which we achieve linear running time in the theory of mind level 
for the given I-POMDP problem (under no environmental uncertainty i.e. the only source of uncertainty being the 
future partner actions).

The components used in the calculation are the observation-action history h $= \{o_0, a_0, o_1, a_1, \ldots \}$ 
(encoding actions taken a and observations made o in temporal sequence, as indicated by a time subindex), 
the target theory of mind level k, the remaining steps 
till the time horizon P, the reward expectation E, the material reward r(a,o, $\theta$) of observing o after action a, the probability p(o) of 
 observing $o$ upon taking action $a$ at history $h$ for a given intentional model $\theta$, the utility of future steps U and importantly the level $-1$ ``default'' choice 
 making model/policy $ \pi_{\textrm{Default}}$, that starts the hierarchy. Furthermore, let $|$A$($h$)|$ be the number of possible 
 actions at history $h$, $ |O( \{$ h, a $\})|$ be the number of possible observations after taking action a at history h, 
 U(o) be the utility of observation $o$ to the agent and $|$E$|$ be the number of elements of the vector E. Additionally, 
 $\Theta (h)$ denotes the belief state of the agent at history h, the probability distributions on the possible partner types, with $\theta$ 
 denoting a concrete intentional model and $|\Theta |$ denoting the number of intentional models that an agent holds and 
 $p(\theta, \Theta (h))$ denotes the probability of $\theta$ under the given Belief state $\Theta (h)$. 
 $\theta `$ denotes the intentional model that the agent themselves is using (in this concrete case, their utility and irritability) 
 and $\pi_{\iota}$ denotes the irritation policy, with $\nu_{\iota}^{\theta}$ denoting the irritation weight under a given current model.

The target of the algorithm \ref{ToMAlg} (procedure ToM-HIERARCHY) is to calculate, for a given number of future steps P,  all encountered histories h 
and ToM-levels k, the 
action probabilities $\mathbb P$(h,P,k, $\theta'$, $\Theta (h)$)  for all possible actions. This is accomplished by means of 
action values/conditional reward expectations Q(h, P, k,$\theta `$, $\Theta (h)$)  obtained by means of dynamic programming/the 
Bellman equation (procedure BELLMAN). Probabilities are assumed to be obtained via a logistic softmax with the action values as 
input (procedure PROB). The values crucially depend on the probabilities of partner actions (at level k-1) $\mathbb P ( \{ $h , a $\}$, P, k$-1,\theta `, \Theta (h))$ 
(procedure OBSERVATION). ``Partner-Irritation'' denotes the current irritation state of the partner (as by the mechanism in the main text) under a given intentional model.

\begin{algorithm}
\caption{Theory of Mind Calculation}\label{ToMAlg} 
\begin{subalgorithm}{.5\textwidth}
\begin{algorithmic}

\Procedure{ToM-HIERARCHY}{h , k, P, $\theta'$, $\Theta (h)$}
	\For {LEVEL $= 0,\ldots,$ k}  
		\If{ mod$($ LEVEL$-$k$, 2)=0$}
			\State Q(h, P, LEVEL, $\theta `$, $\Theta (h)$) =  BELLMAN(h, P, LEVEL, $\theta `$, $\Theta (h)$)	
			\State  $\mathbb P$(h,P, LEVEL, $\theta `$, $\Theta (h)$) $=$  PROB(Q(h, P, LEVEL, $\theta '$, $\Theta (h)$))	
		\EndIf 
	\EndFor

\State \Return $\mathbb P( $h$,$ P$,$k,$\Theta (h))$
\EndProcedure
\Procedure{OBSERVATION}{h,a,P,k, $\Theta (h)$}
	\If{ P $< 0$}
		\State \Return 0
	\EndIf
	\For {o $= 1,\ldots, |O( \{$ h, a $\})|$}
	\State  p(o)$\gets 0$
	\EndFor
	\For {$\theta = 0, \ldots, |\Theta (\{h, a\}) |$ }
	\If{k $> -1$}
	\State $\nu_{\iota}^{\theta}\gets $ Partner-Irritation
	\State Q($\{ $h, a $\}$, P, k $-1$, $\theta$, $\Theta (\{h, a\})$) =\\  BELLMAN($\{$ h, a $\}$, P, k-1, $\theta$, $\Theta (\{h, a\})$)	
	\State  $\mathbb P( \{ $ h, a $\}$, P,k $-1, \theta$, $\Theta (\{h, a\})) =$  \\
	PROB( Q($\{ $h, a $\}$, P, k $-1$, $\theta$, $\Theta (\{h, a\})$))		
	\State  $\mathbb P( \{ $ h, a $\}$, P,k $-1, \theta$, $\Theta (\{h, a\})) =$  \\
	$(1-\nu_{\iota}^{\theta})$  $\mathbb P( \{ $ h, a $\}$, P,k $-1, \theta$, $\Theta (\{h, a\}))$ + $\nu_{\iota}^{\theta}$ $\pi_{\iota}$
	\For {o $= 1,\ldots, |O( \{$ h, a $\})|$}
	
	\State p(o) $ \gets   p(o) + p(\theta, \Theta (\{h, a\})) \mathbb P ( \{ $h , a $\}$, P, k$-1$, $\theta$, $\Theta (\{h, a\})$)(o)
	\EndFor	
	\Else
	\For { o $= 1,\ldots, |O( \{$ h, a $\})|$}
		\State p(o) $ \gets p(o)+ p(\theta, \Theta (\{h, a\}))\pi_{\textrm{Default}} $(o)
	\EndFor

	\EndIf
	\State\Return p		
	\EndFor
\EndProcedure
\end{algorithmic}
\end{subalgorithm}%
\begin{subalgorithm}{.5\textwidth}
\begin{algorithmic}
\Procedure{BELLMAN}{h, P, k, $\theta$, $\Theta (h)$}
	\If{P $< 0$}
		\For {a $= 1,\ldots, |A($h$)|$}
			\State E(a)$ \gets 0$ 
		\EndFor
		\State \Return E
	\EndIf
	
		\For {a $= 1,\ldots, |A($h$)|$}
			\State E(a) $\gets 0$
			\State p = OBSERVATION(h, a, P, k, $\Theta (h)$)
			\For {o $= 1,\ldots, |O( \{$ h, a $\})|$}
				\State h$' \gets \{$h,a,o$\}$  
				\State E(a) $=$ E(a) $+$ r(a,o, $\theta$) p(o) 
				\State U(o) $=$ BELLMAN(h$'$, P$-1$, k, $\theta$ , $\Theta (h')$)			
 			\EndFor
 			\State $\mathbb P$ =PROB(U)
 			\For {o $= 1,\ldots, |O( \{$ h, a $\})|$}			
 			\State E(a) $=$ E(a) $+$ U(o)$\mathbb P$(o)
  			\EndFor			
 		\EndFor
	\State \Return E 
	
\EndProcedure
\\
\Procedure{PROB}{E}
	\State $Sum \gets 0$
	\For {a $= 1,\ldots,$ |E|}
		\State $Sum = Sum + e^{\textrm{E(a)}}$	
	\EndFor
	\For {a $= 1,\ldots,$ |E|}
		\State $\mathbb P(a) = \frac{e^{\textrm{E(a)}}}{Sum}$
	\EndFor
\State \Return $\mathbb P$
\EndProcedure
\end{algorithmic}
\end{subalgorithm}
\end{algorithm}

The algorithm starts from levels $\geq 0$, if they are an even number of levels apart from the target level $k$. The reason for this is, 
that an agent at level $k$ models their partner at level $k-1$ and this partner in turn models the agent at level $k-2$. 

The well known Bellman equation allows to calculate current action preferences based on potential future outcomes, which 
in our case crucially depend on the choice preferences of the partner  Q($\{ $h, a $\}$, P, k $-1$) and the resulting 
likelihood $\mathbb P ( \{ $h , a $\}$, P, k$-1)$ of choosing a response (which is included in the observation o) to the action 
a of the agent. So in one Bellman calculation, both the choice preference of the agent level $k$ and the partner model 
at $k-1$ are being calculated. 

Essentially now the partner model at $k-1$ depends only on the choice preferences of the $k-2$ (and lower) agent models, which at 
that point have been fully calculated and stored. In turn the level $k$ agent can be fully calculated from the level $k-1$ response preferences 
alone, leading to the stated linear complexity in $k$. 

The principles behind this algorithm are not limited to exact calculations (which may be forbiddingly expensive in larger problems). 
Instead the procedure could be combined with approximate solutions methods if one utilizes a ``convergence criterion'' 
for each levels' calculations, before moving to the next ``level-layer''. One potential complication to be aware of in this case, 
is that higher level simulations may choose very different action paths, therefore necessitating a return to lower level calculations, 
if the resulting history was not sufficiently explored at the lower level. This may incur additional costs, compared to a pure 
linear increase in running time as is the case for the exact calculation. 

The various critical scalings are shown in
figure~\ref{fig:alg}A-D:
Figure~\ref{fig:alg}A;B show the linear growth of computation time and
memory respectively with (maximal) theory of mind level (including all
lower-level calculations and tree storage).  Conversely,
figure~\ref{fig:alg}C shows the exponential rise in time for
calculating a level $k^I=4$ investor, level $k^T=3$ trustee interaction
at different planning horizons. Similarly, the exponential growth of
memory use in the planning horizon for a level $k^I=4$ investor, level
$k^T=3$ trustee can be seen in figure~\ref{fig:alg}D.

\begin{center}
\includegraphics[width=5in, height = 5in]{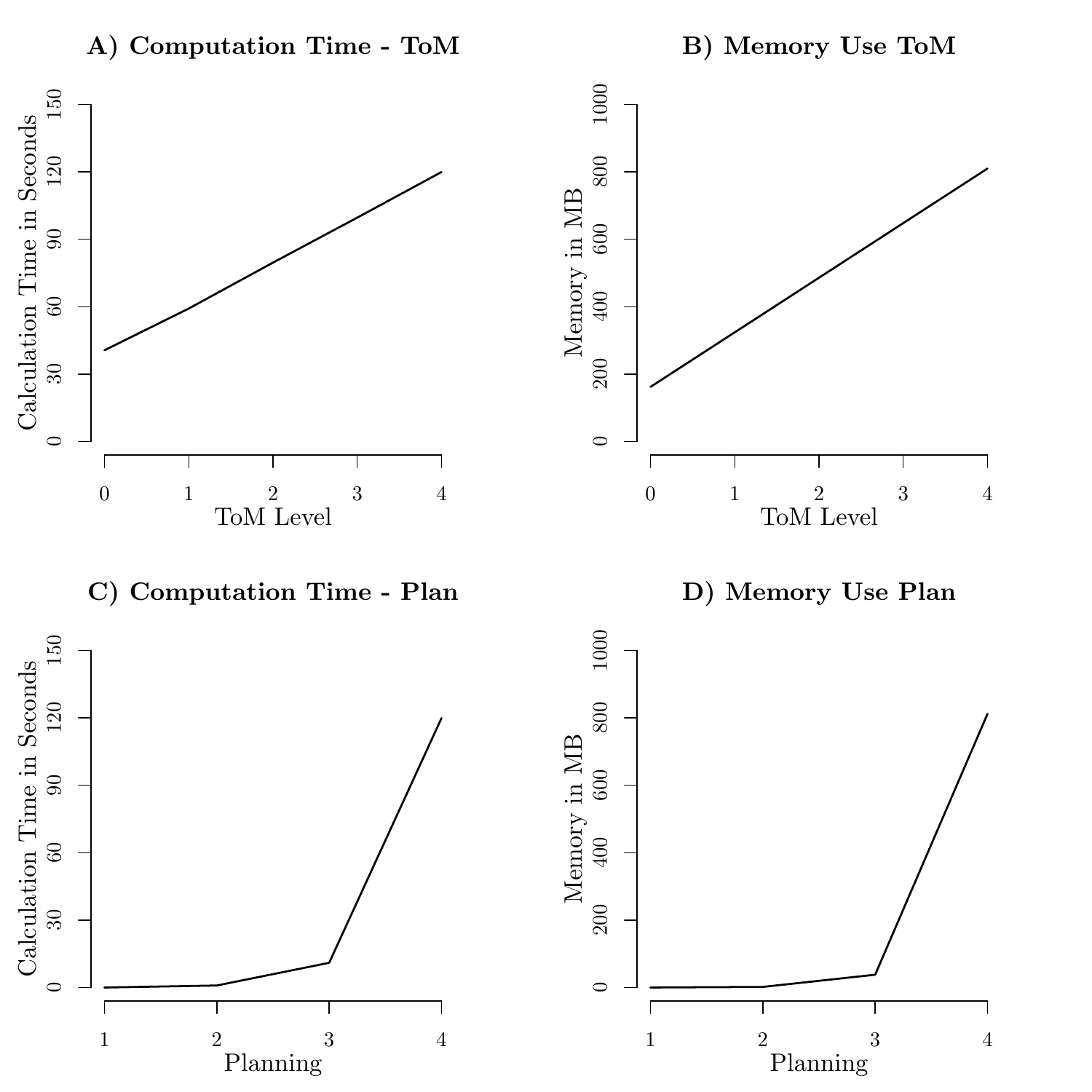}
	\captionof{figure}{\small{ A: Linear scaling of calculation time  with the maximum ToM level 
            used, at planning horizon $4$. 
            B: Linear scaling of memory use with the maximum ToM level used, at 
            planning horizon $4$.  C: Exponential scaling
            of the computation time for a $10$ step
            simulated interaction at ToM $4$ with the planning horizon. 
             D:  Exponential scaling
            of the memory use for a $10$ step
            simulated interaction at ToM $4$ with the planning horizon.}
	\label{fig:alg}}
\end{center}

\end{section}

\end{document}